\documentclass[lettersize,journal]{IEEEtran}
\IEEEoverridecommandlockouts

\usepackage{cite}
\usepackage{amsmath,amssymb,amsfonts}
\usepackage{algorithmic}
\usepackage{graphicx}
\usepackage{textcomp}
\usepackage{subfig}

\usepackage{xcolor}
\usepackage{tabularx}
\usepackage{booktabs}
\usepackage{siunitx}
\usepackage[colorlinks,
            linkcolor=black,
            urlcolor=black,
            anchorcolor=black,            
            citecolor=black]{hyperref}

\usepackage{booktabs} 
\usepackage{multirow} 
\usepackage{balance}
\usepackage{makecell}

\def\BibTeX{{\rm B\kern-.05em{\sc i\kern-.025em b}\kern-.08em
    T\kern-.1667em\lower.7ex\hbox{E}\kern-.125emX}}

\newcolumntype{P}[1]{>{\centering\hspace{0pt}}p{#1}}
\newcolumntype{Z}{>{\centering\let\newline\\\arraybackslash\hspace{0pt}}X}

\makeatletter
\newcommand{\thickhline}{%
    \noalign {\ifnum 0=`}\fi \hrule height 1pt
    \futurelet \reserved@a \@xhline
}
\makeatother

\begin{document}

\title{
SemCovert: Secure and Covert Video Transmission via Deep Semantic-Level Hiding

}



\author{
Zhihan Cao$^1$,
Xiao Yang$^1$,
Gaolei Li$^1$,
Jun Wu$^1$,
Jianhua Li$^1$, and
Yuchen Liu$^2$
\\ \emph{$^1$Shanghai Jiao Tong University, Shanghai, China}
\\ \emph{$^2$North Carolina State University, Chapel Hill, NC, USA}
}





\maketitle

\begin{abstract}
Video semantic communication, praised for its transmission efficiency, still faces critical challenges related to privacy leakage. Traditional security techniques like steganography and encryption are challenging to apply since they are not inherently robust against semantic-level transformations and abstractions. Moreover, the temporal continuity of video enables frame-wise statistical modeling over extended periods, which increases the risk of exposing distributional anomalies and reconstructing hidden content. To address these challenges, we propose SemCovert, a deep semantic-level hiding framework for secure and covert video transmission. SemCovert introduces a pair of co-designed models, namely the semantic hiding model and the secret semantic extractor, which are seamlessly integrated into the semantic communication pipeline. This design enables authorized receivers to reliably recover hidden information, while keeping it imperceptible to regular users. To further improve resistance to analysis, we introduce a randomized semantic hiding strategy, which breaks the determinism of embedding and introduces unpredictable distribution patterns. 
The experimental results demonstrate that SemCovert effectively mitigates potential eavesdropping and detection risks while reliably concealing secret videos during transmission.
Meanwhile, video quality suffers only minor degradation, preserving transmission fidelity. These results confirm SemCovert’s effectiveness in enabling secure and covert transmission without compromising semantic communication performance. 
\end{abstract}


\section{Introduction}
With the rapid evolution of informatic technology and network infrastructure, video data has found extensive deployments across various industries, becoming a pivotal medium for information transmission. Traditional video transmission systems prioritize the complete transmission of raw pixel data. However, under conditions of bandwidth limitations or network instability, large data sizes often result in high transmission delays and data loss, making it difficult to meet real-time demands. Video Semantic Communication (SemCom), as an emerging paradigm, breaks this limitation by extracting and transmitting key semantic information related to the task (instead of raw bits or signals), significantly reducing the data volume and bandwidth consumption, consequently achieving more efficient and reliable transmission in complex network environments. The technology has already proven its practical value in fields such as video conferencing~\cite{9955991}, traffic monitoring~\cite{10974507}, and video question answering~\cite{guo2025videoqa}.

However, despite reducing data volume by simplifying transmission content from raw pixels to semantic information, video semantic communication still faces critical security concerns~\cite{yang2024secure}. While semantic information is compressed, it may still carry sensitive private data. If exposed to illegal surveillance or tampering, this data could result in privacy breaches or lead to misinformed decisions. To address these issues, traditional video transmission security technologies primarily rely on \textit{steganography} and \textit{encryption} techniques~\cite{abomhara2010overview}. Steganography embeds secret information into video by manipulating pixel, frequency, or temporal characteristics. Recently, deep learning-based research has significantly improved steganographic approaches. 
For example, the Convolutional Neural Network (CNN)-based video steganography method in~\cite{telli2024new} improves concealment and robustness, while~\cite{xu2022end} uses Generative Adversarial Networks (GANs) to enhance tamper resistance. Encryption techniques like Advanced Encryption Standard (AES) and Rivest–Shamir–Adleman (RSA), and content-aware encryption are also widely used.
For instance, the content-aware video encryption scheme proposed in~\cite{ShengFLCWS25} adjusts encryption strength according to the video's complexity, thus augmenting both encryption efficiency and security. These technologies protect video data from leakage or tampering.


Nonetheless, traditional methods hold inherent limitations in semantic communication, which focuses on high-level semantic representations rather than raw data. Techniques like steganography and encryption operate at the raw data level and lack the semantic awareness required to preserve embedded secrets, causing them to be misclassified as noise during semantic encoding~\cite{huo2024image}. For example, video steganography designed for text concealment works at the pixel or frequency domain, mismatching with semantic encoders and often leading to the exclusion of secret information.
Moreover, the long-term, continuous semantic streams typical of video semantic communication expand the window for adversarial analysis. By leveraging temporal modeling, attackers can perform cross-frame statistical analyses on intercepted semantic vectors to detect distribution shifts and abnormal temporal patterns caused by embedded data, which aids in inferring embedding mechanisms and reconstructing hidden content~\cite{kunhoth2023video}. Illustratively, attackers may persistently monitor the statistical properties of semantic representations to build distribution-shift-based detection models, accordingly facilitating reverse modeling and recovery of the embedded information.
\textit{Thus, how can we effectively achieve secure and covert video transmission in semantic communication?}

To this end, we propose \textbf{\textit{SemCovert}}, a deep semantic-level hiding framework for secure and covert video transmission. The core idea is straightforward: \textit{secret information is embedded and transmitted within the semantic encoding space so that it remains indistinguishable while coexisting with normal semantic content}. The framework consists of two collaborative components. On the private transmitter side, a semantic hiding model embeds the semantic features of the secret video into the semantic representation of the cover video, generating a single output that appears visually normal and remains fully compatible with standard decoders. As a result, unauthorized receivers can only reconstruct the cover content and remain unaware of any hidden information. To further fortify security, a randomized semantic hiding strategy is introduced to probabilistically determine whether embedding should be performed at a given time step, increasing resistance against detection and analysis. On the authorized receiver side, a secret semantic extractor is deployed to disentangle the hidden semantic branch from the shared data stream and accurately recover the embedded secret content.

\textit{To the best of our knowledge, we are the first to perform semantic-level hiding in video semantic communication}.

In summary, this paper makes the following contributions:

\begin{itemize}
    \item[$\circ$] We design a semantic hiding model and a secret semantic extractor that collaboratively enable seamless embedding of secret information at the semantic level, without compromising the efficiency of semantic transmission. This design ensures that semantic hiding can be naturally integrated into the existing semantic communication pipeline.
    \item[$\circ$] To enhance security and stealthiness, we introduce a randomized semantic hiding strategy and adopt a joint training scheme. This allows the system to maintain reliable semantic transmission while significantly increasing resistance to  attacks, making the hiding behavior difficult to detect or model for adversaries.
    \item[$\circ$] Extensive experiments on multi-resolution and multi-scene video datasets validate the effectiveness of SemCovert: it effectively mitigates potential eavesdropping and detection risks while reliably concealing secret videos during transmission, and incurs only minor degradation in cover video quality, thereby demonstrating its ability to preserve transmission fidelity while enabling accurate and covert information hiding.
\end{itemize}

\begin{table}[!t]
    \centering
    \caption{Secret extraction efficacy of secure transmission methods under SemCom pipeline.}

    \renewcommand{\arraystretch}{1.2} 
    \setlength{\tabcolsep}{6pt}        

    \begin{tabularx}{\linewidth}{ccccc}
    \thickhline
        \textbf{Types} & \textbf{Methods}& \textbf{PSNR$\uparrow$}& \textbf{SSIM$\uparrow$} & \textbf{FVD$\downarrow$}  \\ 
    \midrule
    \midrule
        \multirow{5}{*}{Cover Videos} &
        RoGVS \cite{mao2024covert} & $28.76$ & $0.896$ & $1.474$  \\ 
        & LF-VSN \cite{lvni} & $28.66$ & $0.901$ & $1.572$  \\ 
        & RIS \cite{9878477} & $27.91$ & $0.884$ & $1.924$  \\ 
        & SemCovert (\textit{Ours}) & $28.04$ & $0.891$ & $1.724$  \\ 
        & SemCom (\textit{Baseline}) & $30.69$ & $0.911$ & $0.932$  \\ 

    \midrule
        \multirow{5}{*}{Secret Videos} &
        RoGVS \cite{mao2024covert} & $17.06$ & $0.546$ & $55.94$  \\ 
        & LF-VSN \cite{lvni} & $16.61$ & $0.511$ & $59.57$  \\ 
        & RIS \cite{9878477} & $15.23$ & $0.451$ & $65.79$  \\ 
        & SemCovert (\textit{Ours}) & $27.66$ & $0.881$ & $2.772$  \\ 
        & SemCom (\textit{Baseline}) & $30.32$ & $0.922$ & $0.829$  \\ 
    \thickhline
    
    \label{table.motiva}
    \end{tabularx}
\end{table}

\begin{figure}
    \centering
    \includegraphics[width=0.95\linewidth]{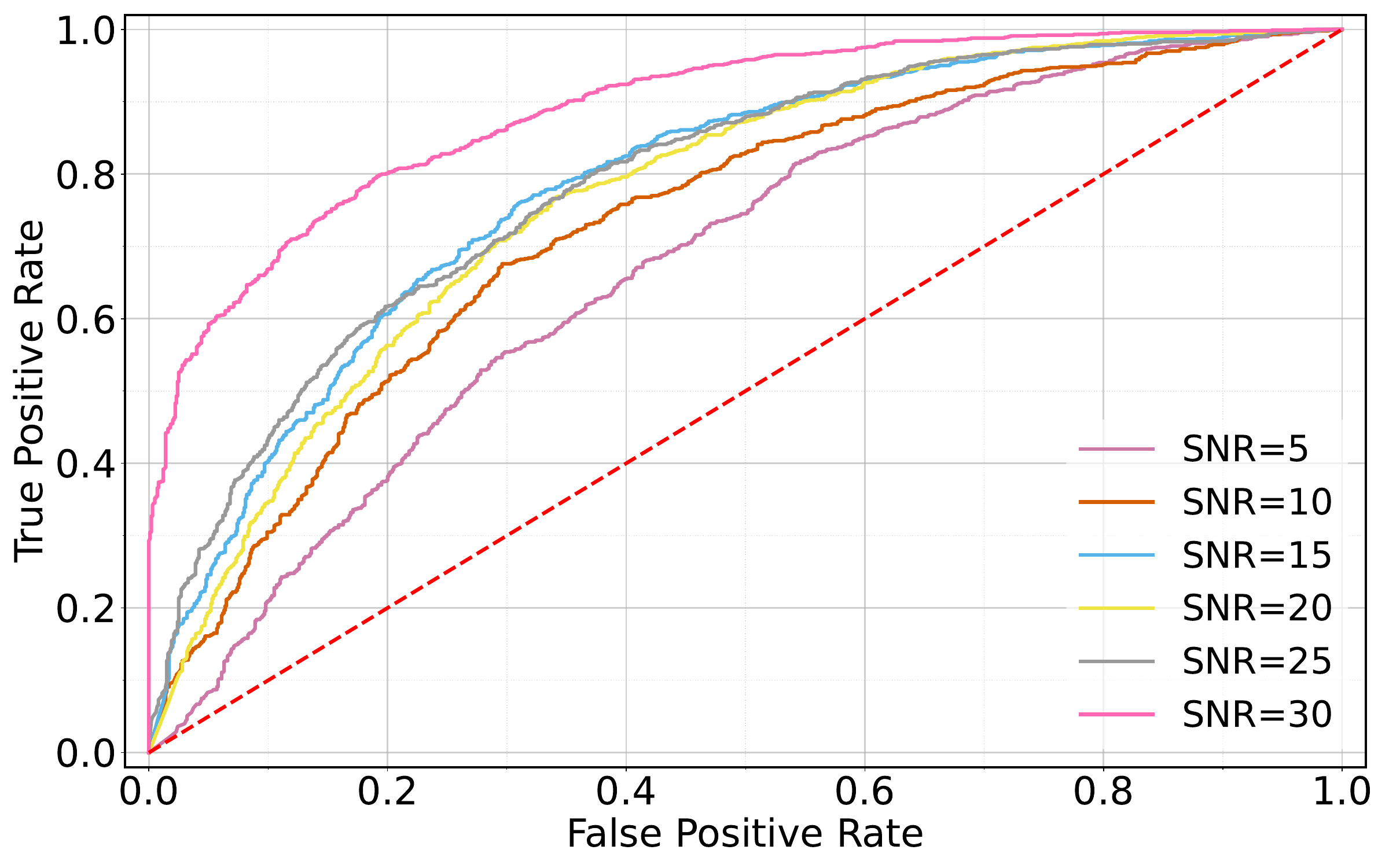}
    \caption{The detection results of potential eavesdroppers regarding the presence of secret transmission. }
    \label{fig:motiva}
\end{figure}

\section{Motivation}
\subsection{Current Limitation}

Current secure transmission approaches face two critical limitations in SemCom scenarios: \textit{\textbf{i}}) they predominantly embed security mechanisms as subtle perturbations in the signal or bit domains, which \textit{fail to preserve availability when semantic representations serve as the transmission unit}; and \textit{\textbf{ii}}) the inherent temporal continuity of semantic video streams \textit{exposes long-range statistical dependencies that adversaries can exploit}, enabling cross-frame statistical analysis to detect distributional shifts and consequently infer or reconstruct the concealed semantic content.
The experimental verification was conducted and is presented in Sec. \ref{pre-exp}.

\subsection{Empirical Verification}
\label{pre-exp}
\textbf{Existing secure transmission schemes exhibit limited effectiveness under SemCom pipeline.} 
To substantiate whether previous methods' efficacy under SemCom paradigm, experiments on the video SemCom architecture were implemented utilizing UCF101 \cite{soomro2012ucf101} dataset. 
With Signal-to-Noise Ratio (SNR) 25\,dB, we examined AWGN (Additive White Gaussian Noise)  via three secure transmission schemes: RoGVS (steganography, \cite{mao2024covert}), LF-VSN (steganography, \cite{lvni}), and RIS (steganography, \cite{9878477}), and reported the Peak Signal-to-Noise Ratio (PSNR), Structural Similarity Index Measure (SSIM), and Frechet Video Distance (FVD).

As demonstrated by the experimental results in Tab.~\ref{table.motiva}. It can be observed that three state-of-the-art and robust steganographic methods are all significantly affected when deployed in a semantic communication framework.
For the cover video, semantic communication is able to preserve task-relevant semantic information during transmission, resulting in relatively stable reconstruction quality. Consequently, the three methods incur only minor performance degradation, with PSNR drops of less than $2$ and negligible variations in SSIM and FVD.
In contrast, the impact on the secret video is much more severe. This is primarily because semantic communication does not treat secret-related information as valid semantics. Since these existing methods embed secret information in the pixel or frequency domain, such information is largely discarded or distorted during semantic compression. Resultantly, the PSNR of the recovered secret video drops to around $16$, SSIM decreases to approximately $0.5$, and FVD increases to about $40$, showing substantial degradation in reconstruction quality and even loss of practical usability.
SemCovert takes into account optimization at semantic level, making its effectiveness remarkably superior to other solutions.
These findings clearly suggest that security mechanisms designed for the pixel or frequency domain are ill-suited for semantic communication, highlighting the necessity of semantic-level approaches for secure and covert transmission.

\vspace{0.8ex}
\textbf{The long temporal structure of video data makes semantic communication more susceptible to detection and analysis.}
We consider a threat model in which an adversary can continuously monitor the semantic communication stream and has access to the semantic encoder–decoder architecture, enabling the training of a dedicated detection model. Under this setting, we adopt R3D~\cite{tran2018closer} as the backbone network to train a binary classifier that determines whether SemCovert is transmitting secret information when the Randomized Semantic Hiding Strategy is not applied. 

Fig.~\ref{fig:motiva} presents the Receiver Operating Characteristic (ROC) curve obtained using the detection model as the detector. The ROC curve characterizes the relationship between the true positive rate (TPR) and the false positive rate (FPR) across different decision thresholds and is widely used to evaluate the discriminative capability of binary classifiers. In the ideal case, when the detector fails to distinguish between normal and secret semantics, its performance degenerates to random guessing, corresponding to an accuracy of approximately $50\%$. Therefore, a ROC curve closer to this ideal condition (red dashed line) indicates higher security and stealthiness.
As observed from the results, under favorable channel conditions (\textit{e.g.}, $\text{SNR} = 30$), the eavesdropper can continuously obtain high-quality semantic representations and train a detector with strong discriminative capability. Only when the channel quality degrades ($\text{SNR} < 10$) does channel noise significantly impair detection performance. These observations indicate that relying solely on channel noise is insufficient for robust security, highlighting the necessity of introducing deliberate confusion mechanisms during transmission to make eavesdropping and detection substantially more difficult.


\subsection{Our Intuition}
The failure of existing methods under SemCom scenarios arises from the absence of a dedicated knowledge framework for secret video transmission within system architecture, resulting in low-efficiency encoding and decoding. 
To resolve this, we can implant secret transmission knowledge into the SemCom system via specified learning procedure alongside regular data transmission knowledge. Concretely, \textit{secret decoding capabilities are deployed exclusively on authorized clients, powering differentiated decoding: \textit{\textbf{i}}) authorized receivers can recover both public and private data, and \textit{\textbf{ii}}) unauthorized ones can solely access public content. }
Additionally, in this paradigm, secret video must be imperceptibly embedded into carrier data during encoding to ensure concealment. From a temporal perspective, \textit{the embedded data should preserve the statistical features of the original carrier to minimize detectability introduced by information hiding}.
This design assures secure transmission of sensitive information and preserves the functionality of the public communication channel.

\section{System Overview}
\begin{figure*}[t]
    \centering
    \includegraphics[scale=0.55]{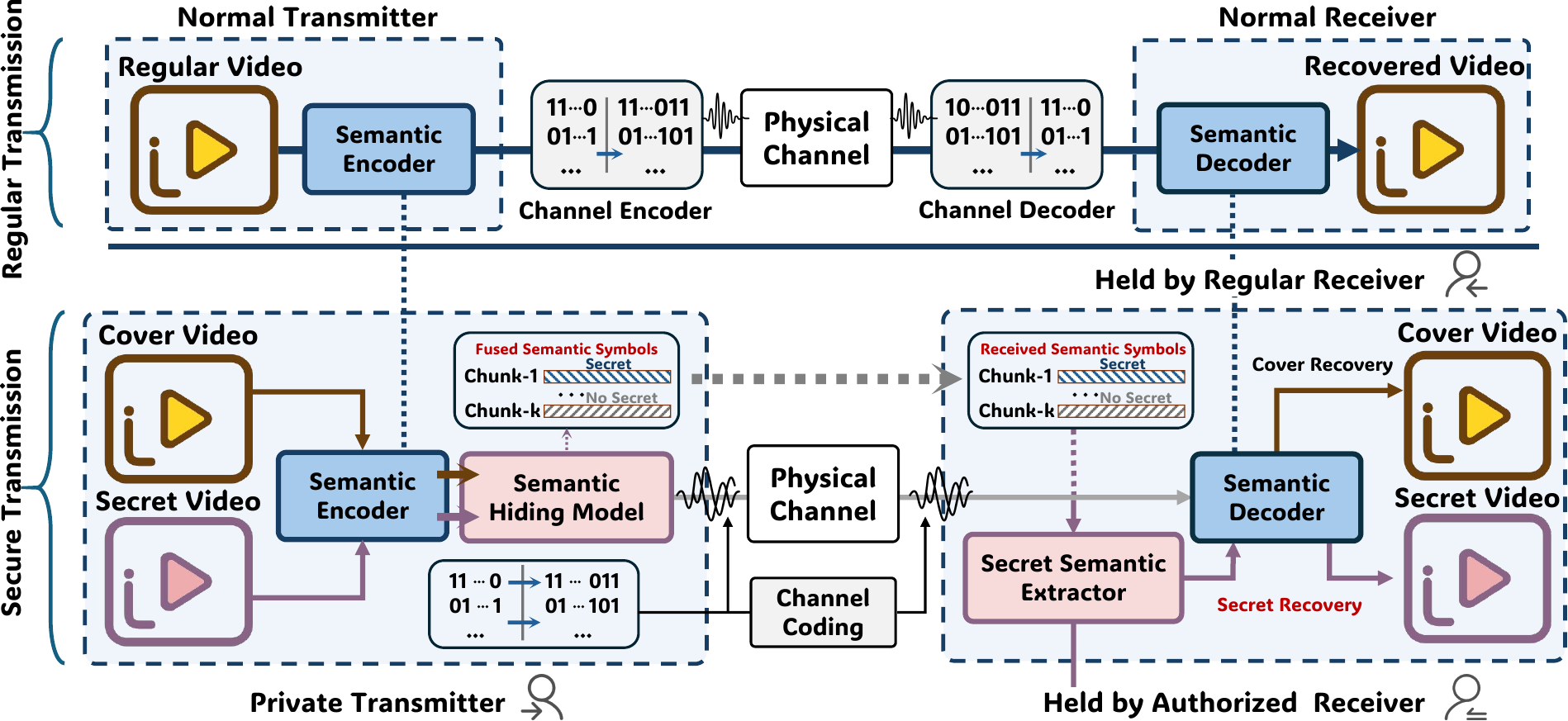} 
    \caption{Illustration of the proposed \textbf{SemCovert}, wherein the transmitter exploits a Semantic Hiding Model to imperceptibly embed secrets videos into cover stream. At the receiving peer, only authorized clients possessing the Secret Semantic Extractor can decode the hidden information, while regular users' Semantic Decoders can solely extract the cover video.}
    \label{fig.overview}
\end{figure*}

As illustrated in Fig.~\ref{fig.overview}, the traditional video semantic communication pipeline (upper part of the figure) consists of the following steps: a regular video is first processed by a shared semantic encoder to extract semantic representations, which are subsequently channel-encoded and transmitted over the physical channel. At the receiver side, the signal undergoes channel decoding, followed by semantic decoding to recover the semantic representations, which are then reconstructed into the final video.

SemCovert (lower part of the figure) builds upon this framework by introducing two key modules to enable enhanced utilization of semantic signals: the \textbf{Semantic Hiding Model} and the \textbf{Secret Semantic Extractor}. 

On the sender side, both the cover video and the secret video are first divided into chunks and then encoded into their respective semantic representations. To further enhance security, we introduce a randomized semantic hiding strategy that probabilistically determines whether to embed secret semantic information within each video chunk.
The Semantic Hiding Model embeds the semantic signal of the secret video into that of the cover video, producing a fused hidden semantic representation. This fused semantic signals are then channel-encoded and transmitted over the physical channel.

At the receiver side, different receivers possess different levels of decoding capabilities: \textit{\textbf{i}}) the regular receiver, equipped only with a conventional semantic decoder, can recover only the cover video;  \textit{\textbf{ii}}) the authorized receiver, additionally equipped with the Secret Semantic Extractor, can further extract the embedded secret semantic signal from the hidden semantic representation and reconstruct the corresponding secret video using a semantic decoder.

It is worth emphasizing that SemCovert achieves end-to-end secure and covert transmission by inserting only two modules around the semantic encoder and decoder, without altering the core structure of the existing communication system. This design ensures high compatibility and scalability. Moreover, by performing embedding at the semantic level, SemCovert not only preserves the reconstruction quality of the cover video but also significantly enhances the stealthiness and security of the hidden information.

\section{Main Design}
We detail the core design of SemCovert, including the structures and working mechanisms of its key modules, along with the corresponding joint training strategy.

\subsection{Key Modules Design}
\vspace{0.8ex}
\noindent \textbf{Semantic Encoder and Semantic Decoder.} 
The Semantic Encoder is responsible for extracting high-level semantic representations from the input video, while the Semantic Decoder reconstructs the original video data from the semantic features. SemCovert features a modular and scalable design that is independent of specific encoder–decoder architectures. After joint training of the encoder and decoder to ensure alignment, it can be seamlessly integrated into any semantic-aware video transmission system and adapt to diverse scenarios without architectural or contextual constraints. In this work, we adopt Wan-VAE~\cite{wan2025} as the encoder–decoder backbone due to its strong performance across various video generation tasks. Wan-VAE demonstrates stable and efficient modeling of the semantic space, making it well-suited for this role.

\begin{figure}[t]
    \centering
    \includegraphics[width=0.95\linewidth]{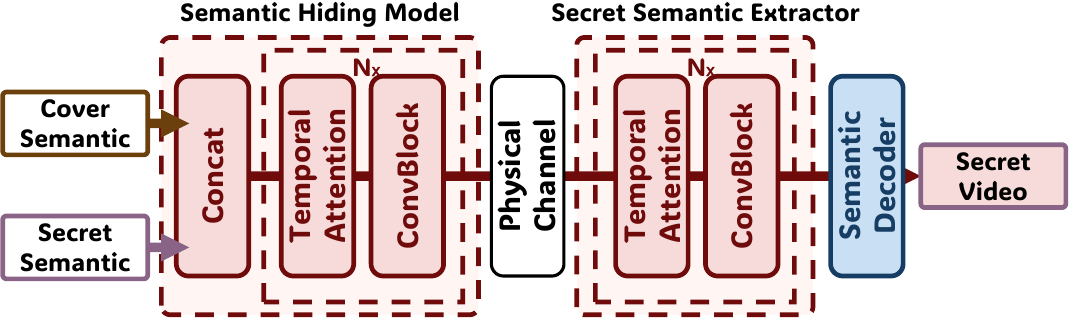}
    \caption{The model architecture of the Semantic Hiding Model and the Secret Semantic Extractor.}
    \label{fig:detail}
\end{figure}

\vspace{0.8ex}
\noindent \textbf{Semantic Hiding Model.} 
As shown in Fig.~\ref{fig:detail}, this module takes the semantic representations of the cover video and the secret video as input. The two semantic features are concatenated along the channel dimension to form a unified feature map. For feature extraction, a multi-head attention mechanism is applied along the temporal dimension to capture long-range dependencies across frames, while stacked 2D convolutional blocks are employed in the spatial dimension to extract local structural patterns. To enhance training stability and accelerate convergence, RMSNorm~\cite{zhang2019rootmeansquarelayer} is applied between consecutive convolutional layers. The final output is compressed along the channel dimension to match the original cover semantics, enabling redundancy-free semantic embedding and ensuring compactness and efficiency. It is made highly similar to the corresponding output without secret embedding, achieving indistinguishability and enhancing security and stealthiness. 

\vspace{0.8ex}
\noindent \textbf{Randomized Semantic Hiding Strategy.}
To enhance resistance against model inversion and statistical analysis, we propose a randomized semantic hiding strategy that intentionally breaks the deterministic nature of conventional embedding. Rather than embedding secret information into every semantic chunk of the cover video, our approach operates on chunked semantic representations and randomly selects a subset of $M$ out of $N$ cover chunks ($M \leq N$) for embedding. This selection exploits the intrinsic redundancy of semantic content to introduce randomness into the embedding process.
Formally, let $I \subset \{1, \dots, N\}$ with $|I| = M$ denote the indices of the selected chunks. Only those chunks at positions $i \in I$ are modified to carry the secret information, while the remaining chunks remain unaltered. 
This results in a spatially sparse and dynamically varying embedding pattern, which substantially increases the difficulty for adversaries to detect or model the embedding behavior from observed input-output correlations. 

Such randomized embedding introduces uncertainty not only for adversaries but also for the decoder, as hidden semantics appear in a non-deterministic manner across chunks. Consequently, the decoder must autonomously determine whether a given semantic chunk contains secret information. Let $\mathbf{z}_i$ denote the semantic representation of the $i$-th chunk; under the proposed strategy, $\mathbf{z}_i$ may be either a normal semantic embedding ($i \notin I$) or a hidden one ($i \in I$), without any explicit indicator. Although sharing the index set $I$ or the random seed between sender and receiver could resolve this ambiguity, it would introduce practical issues such as key management and synchronization. Instead, we formulate decoding as an implicit semantic discrimination problem and enforce this capability during training, enabling the decoder to adaptively identify and extract hidden semantics at runtime without requiring shared randomness.


\vspace{0.8ex}
\noindent \textbf{Secret Semantic Extractor.} 
Also illustrated in Fig.~\ref{fig:detail}, this module takes as input the semantic signal decoded from the physical transmission channel. To preserve consistency in the semantic feature space and maintain architectural symmetry, the Secret Semantic Extractor closely mirrors the structure of the Semantic Hiding Model. Additionally, it must be able to discern whether hidden information is present in the semantic signal: \textit{\textbf{i}}) if the signal contains the semantics of a secret video, the module should accurately extract the hidden semantic content and reconstruct the secret video via the semantic decoder. \textit{\textbf{ii}}) if no secret is embedded, the module should output an invalid placeholder semantic representation that guides the decoder to produce an all-zero video, thereby explicitly indicating that no hidden content was transmitted. This design ensures that the secret video can be reliably and completely recovered, even under the randomized semantic hiding strategy.

\subsection{Joint Training for Randomized Semantic Hiding}

To achieve joint encoding and high-fidelity reconstruction of both cover and secret videos, we propose a comprehensive loss function that integrates reconstruction loss, perceptual similarity, distribution constraints, and VAE-related regularization terms.
We formulate the model training as a multi-objective optimization problem, aiming to preserve the visual quality and semantic integrity of the cover video while enabling high-quality reconstruction of the hidden video. The total loss is
\begin{equation}
\begin{aligned}
    \mathcal{L}_{\text{total}} =\;
    &\lambda_c \cdot \mathcal{L}_{\text{cover}} 
    + \lambda_s \cdot \mathcal{L}_{\text{secret}} 
    + \lambda_p \cdot \mathcal{L}_{\text{perceptual}} \\
    & + \lambda_{\text{kl}}^c \cdot \mathcal{L}_{\text{KL}}^c 
    + \lambda_{\text{kl}}^s \cdot \mathcal{L}_{\text{KL}}^s \\
    &+ \lambda_e \cdot \mathcal{L}_{\text{embedding}} + \lambda_n \cdot \mathcal{L}_{\text{null}}.
\end{aligned}
\end{equation}
The scalar weights $\lambda_*$ balance the contributions of the respective loss components, and the total loss is defined as:

\vspace{0.8ex}
\noindent \textbf{Reconstruction Loss.} 
To enhance robustness, we adopt the Charbonnier loss for measuring reconstruction quality:
\begin{equation}
    \mathcal{L}_{\text{charb}}(x, y) = \frac{1}{N} \sum_{i=1}^{N} \sqrt{(x_i - y_i)^2 + \varepsilon^2},
\end{equation}
where $\varepsilon > 0$ is a small constant to ensure numerical stability. $x$ and $y$ denote predicted and ground-truth video sequences.
We define $
\mathcal{L}_{\text{cover}} = \mathcal{L}_{\text{charb}}(\hat{v}_c, v_c)$ and 
$
\mathcal{L}_{\text{secret}} = \mathcal{L}_{\text{charb}}(\hat{v}_s, v_s)
$,
with $\hat{v}_c, \hat{v}_s$ denoting the reconstructed outputs, and $v_c, v_s$ symbolizing the original inputs.

\vspace{0.8ex}
\noindent \textbf{Perceptual Loss.} 
To address the limitations of pixel-wise metrics, we incorporate perceptual loss computed via the first 16 layers of the VGG-16 network~\cite{simonyan2014very}:
\begin{equation}
    \mathcal{L}_{\text{perceptual}} = \frac{1}{M} \left\| \phi(\hat{v}_c) - \phi(v_c) \right\|_2^2 + \frac{1}{M} \left\| \phi(\hat{v}_s) - \phi(v_s) \right\|_2^2,
\end{equation}
where $\phi(\cdot)$ extracts semantic features and $M$ is the total number of features.

\vspace{0.8ex}
\noindent \textbf{KL (Kullback-Leibler) Divergence Regularization.} 
To regularize the latent space toward a standard normal distribution, we adopt the VAE KL divergence term~\cite{kingma2013auto}:
\begin{equation}
\begin{aligned}
    \mathcal{L}_{\text{KL}} 
    &= D_{\text{KL}}(q(z|\mu, \sigma) \, || \, \mathcal{N}(0, I)) \\
    &= \frac{1}{N} \sum_{i=1}^{N} \left[ \mu_i^2 + \sigma_i^2 - \log \sigma_i^2 - 1 \right],
\end{aligned}
\end{equation}
which is applied to both the cover and secret video encoders, resulting in $\mathcal{L}_{\text{KL}}^c$ and $\mathcal{L}_{\text{KL}}^s$.

\vspace{0.8ex}
\noindent \textbf{Embedding Constraint Loss.} 
To ensure that the fused latent features preserve the distribution characteristics of the original cover features, we introduce a KL-divergence-based constraint:
\begin{equation}
\begin{aligned}
    \mathcal{L}_{\text{embedding}} = \frac{1}{N} \sum_{i=1}^{N} \Bigg[
    \log\left( \frac{\sigma^2_{\text{fused}, i}}{\sigma^2_{\text{cover}, i}} \right) \\
    + \frac{\sigma^2_{\text{cover}, i} + (\mu_{\text{fused}, i} - \mu_{\text{cover}, i})^2}
           {\sigma^2_{\text{fused}, i}} - 1 \Bigg],
\end{aligned}
\end{equation}
where $\mu_*, \sigma^2_*$ are the means and variances from the latent distributions of cover and fused encoders.

\vspace{0.8ex}
\noindent \textbf{Null Secret Loss.} 
Owing to the randomized semantic hiding strategy and channel noise, the extractor may falsely decode a secret from inputs that actually contain no hidden information. To ensure this output contains no meaningful information, we define null secret loss:
\begin{equation}
    \mathcal{L}_{\text{null}} = \mathcal{L}_{\text{charb}}(\hat{v}_s, \mathbf{0}).
\end{equation}
This design encourages the extractor to suppress secret decoding when no hidden content is present, while ensuring that the authorized receiver can still correctly extract the secret information under the randomized semantic hiding strategy.

\vspace{0.8ex}
\noindent \textbf{Mixed Training Strategy}. 
To maintain the performance of the semantic encoder-decoder and to prevent the secret semantic extractor from hallucinating semantic content that was never embedded, we adopt a mixed training strategy. Specifically, the model is trained with two types of samples:
\textit{\textbf{i}}) \textbf{Cover-Secret pairs.} All loss terms are applied, including secret reconstruction loss and embedding constraint, ensuring accurate recovery of hidden semantics and alignment of latent distributions.
\textit{\textbf{ii}}) \textbf{Secret-Free samples.} Secret-related losses are excluded to avoid false learning signals. Instead, a null-secret loss $\mathcal{L}_{\text{null}}$ is applied to explicitly penalize the extraction of any semantic information where no secret has been embedded. This strategy not only preserves the semantic communication system’s original functionality, but also ensures that secret information can be faithfully and selectively extracted under the randomized semantic hiding strategy, thereby enhancing the overall reliability and security of SemCovert.

\section{Evaluation}
\subsection{Experimental Setup}
\noindent \textbf{Datasets Settings.}
To demonstrate the system's performance and assess the impact of the two newly introduced modules on the video semantic communication system, we conducted experiments using the following three video datasets. 

\begin{itemize}
\item[$\circ$] UCF101\cite{soomro2012ucf101}: UCF101 serves as a benchmark dataset for real-world action recognition, encompassing $13,320$ video clips across $101$ distinct action categories, each with a resolution of $320\times240$. To control the data volume while maintaining category diversity, we randomly selected two videos from each class for training and another two for testing. The resulting training and testing sets each include approximately $200$ videos, which are sufficient to capture various video scenes.

\item[$\circ$] DAVIS\cite{Perazzi2016}: DAVIS set contains video sequences designed for video object segmentation. In this study, we utilize the officially provided downsampled version, with each video having a $640\times480$ resolution. It is used solely as a testing set to evaluate the performance of SemCovert in clear and coherent real-world video scenarios.

\item[$\circ$] MOT17\cite{milan2016mot16}: MOT17 is a collection of $15$ pedestrian tracking sequences captured under diverse scenarios, featuring varying camera motions and lighting conditions. We selected the videos with a resolution of $1920\times1080$ as the testing set to evaluate the performance of SemCovert under high-definition conditions representative of real-world deployment environments.
\end{itemize}

\vspace{0.8ex}
\noindent \textbf{Implementation Details.}
All experiments were conducted in a virtualized environment. The configurations utilized for both training and testing are as follows: CUDA 12.8, PyTorch 2.7.1, Python 3.10, and Ubuntu 22.04, running on two NVIDIA RTX 4090 GPUs, each with 24\,GB of memory.

The training and testing are all performed under an assumed AWGN channel model, where the received signal is given by $ y = x + n $, with $ n \sim \mathcal{N}(0, \sigma^2) $.
Following the Wan-VAE output format, channel features of dimension $(16, (T-1)/4 +1, W/8, H/8)$ are flattened and power-normalized to form the transmitted signal. 
To reduce memory consumption and expand the operational space of the randomized semantic hiding strategy, we use $5$ frames as a standard video chunk and use multiple batches to increase speed.
The Semantic Hiding Model and Secret Semantic Extractor use 2D convolutions with kernel sizes $[3, 5, 7]$ for spatial modeling and $8$-head multi-head attention along the temporal dimension for temporal dependency modeling. 
Both networks have a latent dimension of $96$, a stacking depth of $N = 4$, and maintain consistent input-output dimensions.
Additionally, different learning rates are used for training individual modules. 
The semantic encoder-decoder, which has been well pre-trained on other datasets, is fine-tuned with a lower learning rate of $2 \times 10^{-5}$ to adapt to the channel environment. 
In contrast, the Semantic Hiding Model and Secret Semantic Extractor are trained with a higher learning rate of $4 \times 10^{-4}$ to facilitate faster convergence toward their specific objectives.

\subsection{Evaluation Results}

\textbf{SemCovert demonstrates strong stealthiness in  semantic communication.}
Under the joint effect of the embedding constraint loss and the randomized semantic hiding strategy, SemCovert exhibits high semantic-level similarity, making it difficult for attackers to detect or compromise the hidden information. To validate this property, we conducted a comparative analysis between normal semantic signals and those embedded with secret information across three dimensions: \textit{i}) numerical difference (Mean Square Error, MSE), \textit{ii}) directional similarity (Cosine Similarity), and \textit{iii}) distributional divergence (Wasserstein Distance). The results are shown in Tab.~\ref{table.exp1}.
Experimental results across all three datasets show that the MSE between the two types of semantic signals remains below $0.0007$, Cosine Similarity exceeds $0.86$, and the average Wasserstein Distance is less than $0.0085$. Together, these results show that semantic signals carrying secret information closely resemble the original signals in numerical value, directional consistency, and distribution, thereby exhibiting strong concealment in the representation space.
These findings provide strong evidence that SemCovert possesses a high degree of semantic-level stealthiness, enabling the effective embedding of secret information without significantly altering the semantic characteristics of the transmitted content.

\begin{table}[!t]
    \renewcommand{\arraystretch}{1.2}
    \setlength{\tabcolsep}{10pt} 
    \centering
    \caption{Multi-dimensional comparison of semantic similarity between SemCovert and SemCom.}
    \begin{tabularx}{\linewidth}{>{\centering\arraybackslash}m{1.1cm} >{\centering\arraybackslash}X >{\centering\arraybackslash}X >{\centering\arraybackslash}X} 
    \thickhline
        \multirow{2}{*}{\textbf{Dataset}} &
        \textbf{Mean Square Error$\downarrow$} &
        \textbf{Cosine Similarity$\uparrow$} &
        \textbf{Wasserstein Distance$\downarrow$} \\
    \midrule
    \midrule
        \textbf{UCF101} & $0.00071$ & $0.86979$ & $0.00605$ \\
        \textbf{DAVIS}  & $0.00078$ & $0.86034$ & $0.00834$ \\
        \textbf{MOT17}  & $0.00061$ & $0.88016$ & $0.00865$ \\
    \thickhline
    \end{tabularx}
    \label{table.exp1}
\end{table}

\vspace{0.8ex}
\textbf{SemCovert employs the randomized semantic hiding strategy to increase the complexity of detection.} In this experiment, we assume that an attacker’s detector can, in principle, be trained using correctly labeled samples of normal semantics and secret-embedded semantics. However, the introduction of the randomized semantic hiding strategy deliberately mixes normal and secret-carrying semantics during transmission, inevitably injecting label noise into training process and markedly degrading the detector’s learning capability.

\begin{figure}[t]%
    \centering
    \includegraphics[width=0.85\linewidth]{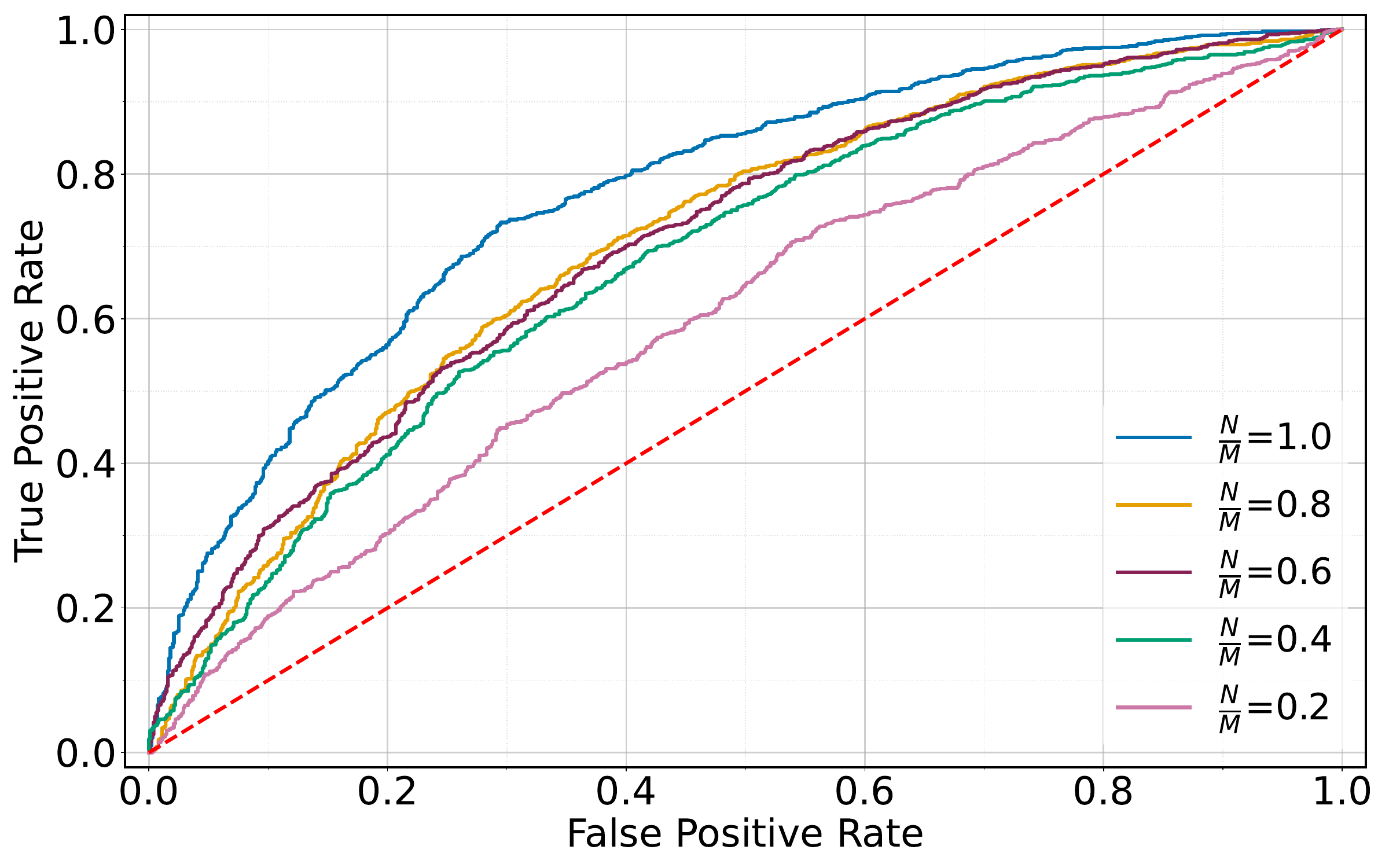}
    \caption{The results of the detection model. The closer the detection accuracy is to $50\,\%$, the higher the security is. $\frac{N}{M}$ represents the capacity ratio employed in the randomized semantic hiding strategy.}
    \label{fig.roc}
\end{figure}

Fig.~\ref{fig.roc} presents the Receiver Operating Characteristic (ROC) curve obtained using an R3D~\cite{tran2018closer} model as the detector. The ROC curve characterizes the relationship between the True Positive Rate (TPR) and the False Positive Rate (FPR), and is widely used to evaluate the discriminative capability of binary classifiers across different decision thresholds. In the ideal case, when the detector fails to distinguish between normal and secret semantics, its detection performance approaches random guessing, corresponding to an accuracy of approximately $50\%$. Therefore, a ROC curve closer to this ideal condition (red line) indicates higher security and stealthiness.
The experimental results demonstrate that the randomized semantic hiding strategy is highly effective across different capacity ratios $\frac{M}{N}$. When the capacity ratio is large, the ROC curve deviates noticeably from the ideal case, indicating that the detector still retains limited discriminative ability. In contrast, when the capacity ratio is reduced , the ROC curve approaches the ideal state, suggesting that the detector fails to learn a reliable discrimination rule. This confirms that the randomized semantic hiding strategy disrupts the deterministic correspondence between semantics and labels, significantly constraining the effectiveness of detection models even when trained by knowledgeable adversaries.

\begin{figure}[t]%
    \centering
    \includegraphics[width=0.85\linewidth]{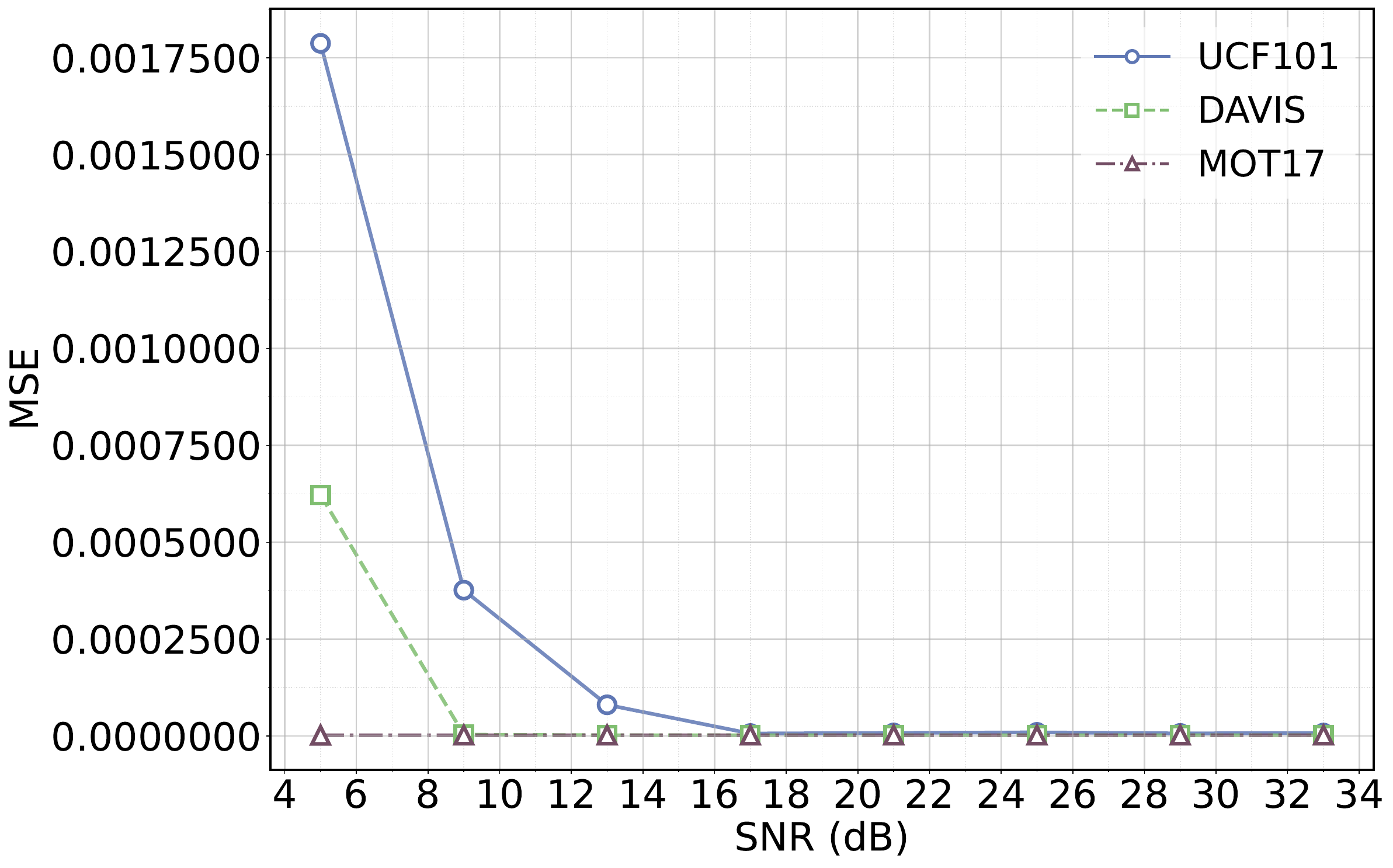}
    \caption{SemCovert's ability to distinguish secret and non-secret carriers from semantic streams. Lower MSE relative to all-zero video indicates better discernment of non-secret cases.}
    \label{fig.mse}
\end{figure}

\vspace{0.8ex}
\textbf{SemCovert still performs well under the randomized semantic hiding strategy.}
This is illustrated intuitively in Fig.~\ref{fig.mse}. During the system design phase, we explicitly defined the no-secret condition to decode into an all-zero video, ensuring that the semantic extractor produces no unintended content when no secret is embedded.

To assess the effectiveness of the design, we measured the Mean Square Error (MSE) between the semantic extractor’s output without secret transmission and its response to an all-zero video. On the UCF101 dataset, even under low-SNR conditions where noise may introduce minor decoding artifacts, the resulting MSE remains at the very low level of \num{e-3} indicating a minimal degree of false extraction. As channel conditions improve (i.e., higher SNR), the MSE rapidly drops to around \num{e-8}, showing virtually no difference from the reference, and confirming that under good channel conditions, the system almost never produces hallucinated outputs. On the MOT17 dataset, SemCovert exhibits even greater robustness. No noticeable hallucinations were observed even under high noise levels, with MSE consistently approaching zero. These results provide strong evidence that SemCovert possesses excellent discriminative capability: it can clearly distinguish between normal semantic transmission and covert semantic transmission scenarios, reliably extract secret information when present, and effectively suppress false decoding when no secret exists—thereby guaranteeing both the security and accuracy of the system.

\vspace{0.8ex}
\textbf{SemCovert effectively prevents the leakage of secret information in the reconstructed cover videos.}
Tab.~\ref{table.exp2} presents the differences between the recovered and original cover videos at three levels: pixel-wise (PSNR), structural (SSIM), and perceptual (FVD). To further assess perceptual-level differences, we compute FVD using three different neural network architectures: FVD1 is based on S3D~\cite{xie2018rethinking}, FVD2 on R3D~\cite{tran2018closer}, and FVD3 utilizes Video Swin Transformer~\cite{liu2022video}.

Experimental results show that across all three datasets, the differences in PSNR, SSIM, and FVD1–FVD3 between the recovered and original cover videos remain minimal, indicating a high degree of consistency across multiple dimensions. In contrast, when comparing the recovered cover videos to the secret videos, the differences are substantial: The PSNR value is usually below $10$, while the FVD score rises significantly, especially in FVD3 (Swin Transformer), where the difference reaches around $200$, indicating a clear perceived gap. These findings confirm that the recovered cover videos do not reveal any information from the secret videos at the pixel, structural, or perceptual levels, thus demonstrating the strong stealthiness and security of SemCovert.

\begin{table}[!t]
    \centering
    \renewcommand{\arraystretch}{1.3}  
    \caption{Semantic and structural similarity analysis in recovered cover videos relative to original covers and secrets.}
    \begin{tabularx}{\linewidth}{P{1.7cm}ZZZZZ}
    \thickhline
            \textbf{Cover Videos}& \textbf{PSNR$\uparrow$}& \textbf{SSIM$\uparrow$} & \textbf{FVD1$\downarrow$} & \textbf{FVD2$\downarrow$}\ & \textbf{FVD3$\downarrow$} \\ 
        \midrule
        \midrule
            \textbf{UCF101} & $26.76$ & $0.886$ & $2.074$ & $4.871$ & $7.329$ \\ 
            \textbf{DAVIS} & $27.66$ & $0.841$ & $4.572$ & $2.547$ & $11.83$ \\ 
            \textbf{MOT17} & $33.66$ & $0.951$ & $0.724$ & $1.211$ & $2.535$ \\ 
        \toprule
        \textbf{Secret Videos}& \textbf{PSNR$\downarrow$}& \textbf{SSIM$\downarrow$} & \textbf{FVD1$\uparrow$} & \textbf{FVD2$\uparrow$}\ & \textbf{FVD3$\uparrow$} \\ 
        \midrule
        \midrule
            \textbf{UCF101} & $8.352$ & $0.205$ & $98.06$ & $79.68$ & $219.3$ \\
            \textbf{DAVIS} & $8.417$ & $0.309$ & $75.84$ & $99.99$ & $225.9$ \\ 
            \textbf{MOT17} & $9.801$ & $0.459$ & $42.46$ & $115.9$ & $186.6$ \\ 
    \thickhline
    
    \label{table.exp2}
    \end{tabularx}
\end{table}

    

\begin{figure*}[t]%
    \centering
    \subfloat[PSNR comparison in UCF101. Size: 240p.]{
        \label{fig.main_result1}
        \includegraphics[width=0.32\linewidth]{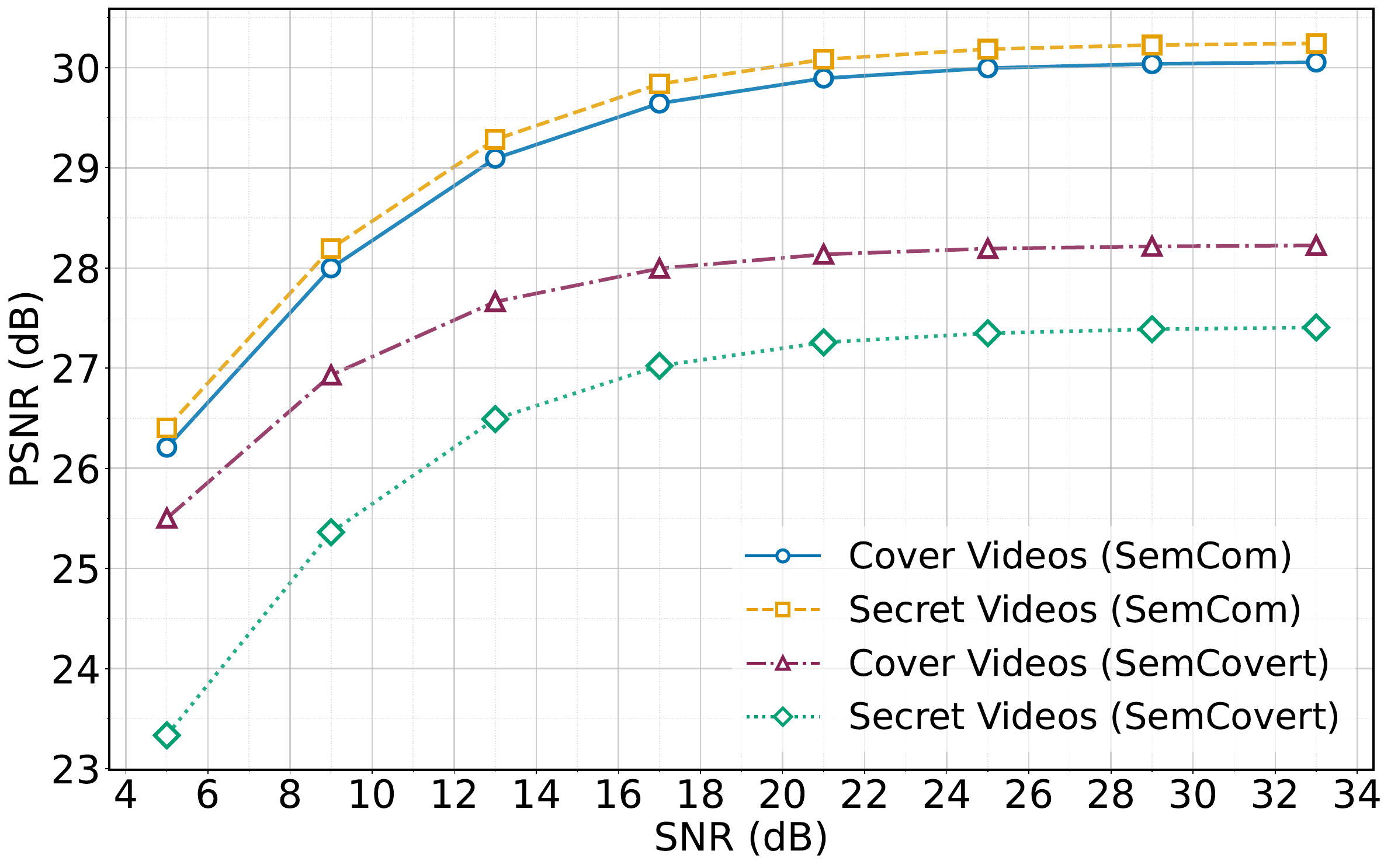}
        }
    \subfloat[SSIM comparison in UCF101. Size: 240p.]{
        \label{fig.main_result2}
        \includegraphics[width=0.32\linewidth]{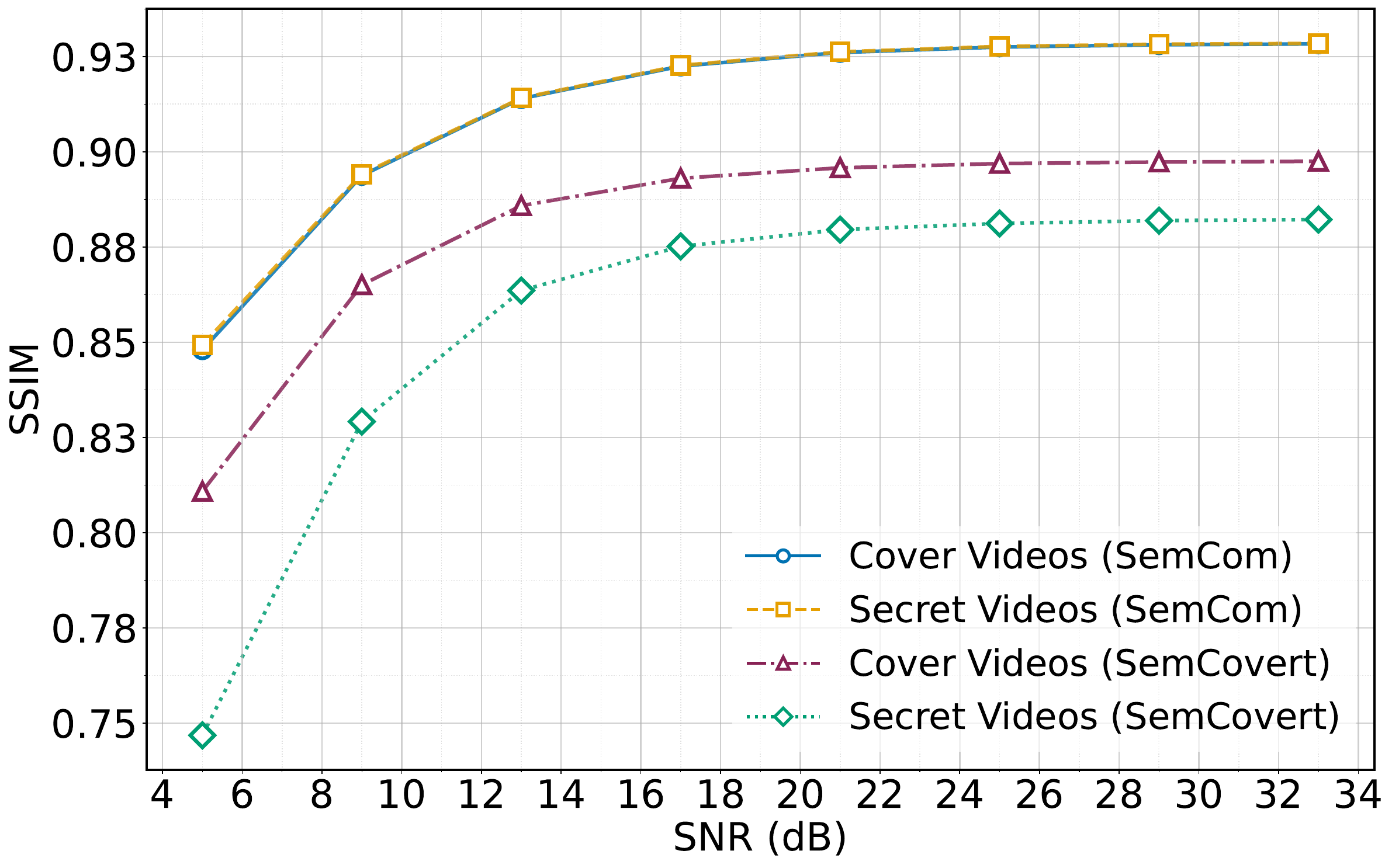}
        }
    \subfloat[FVD comparison in UCF101. Size: 240p.]{
        \label{fig.main_result3}
        \includegraphics[width=0.32\linewidth]{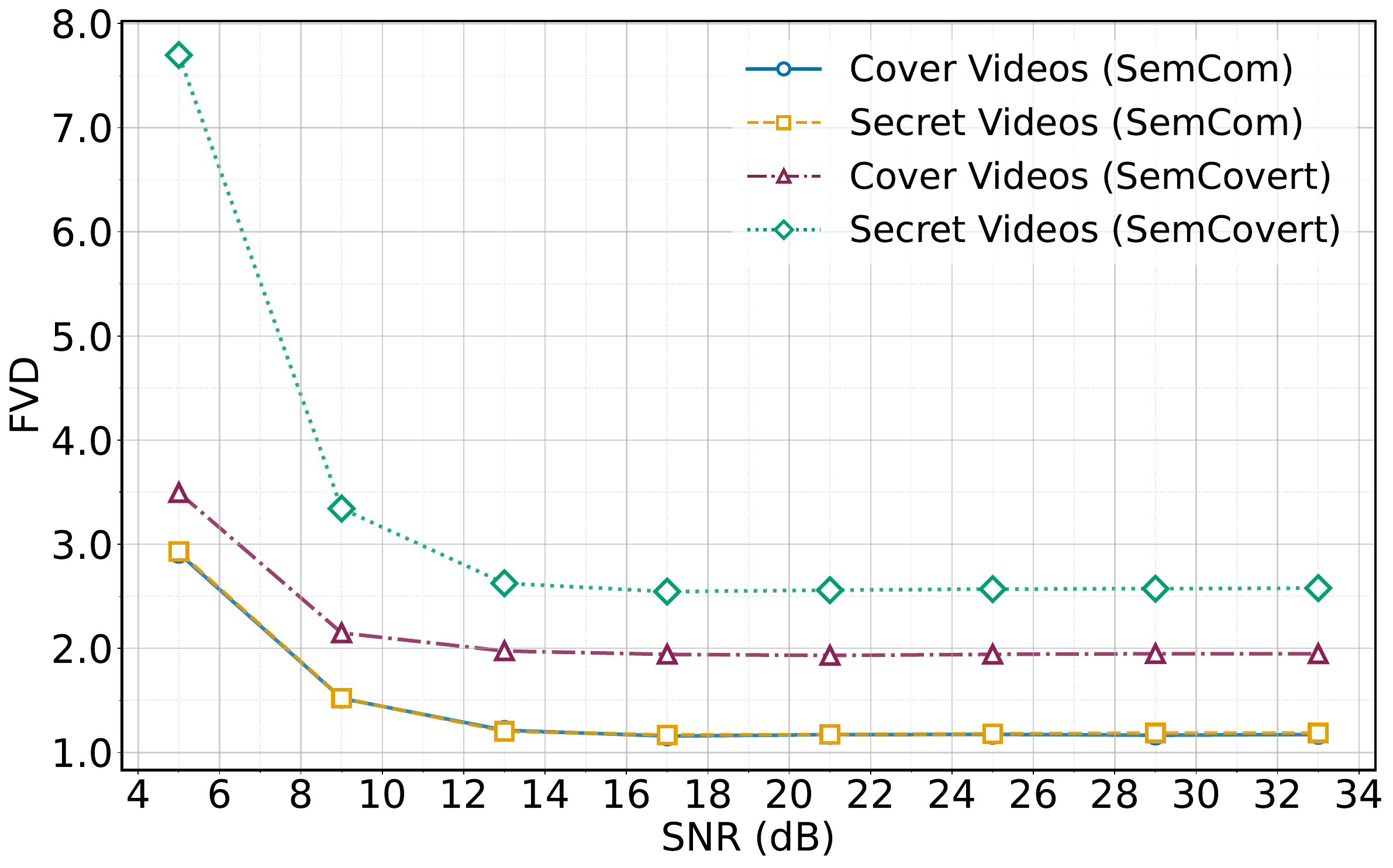}
        }
    \hfill
    \subfloat[PSNR comparison in DAVIS. Size: 480p.]{
        \label{fig.main_result4}
        \includegraphics[width=0.32\linewidth]{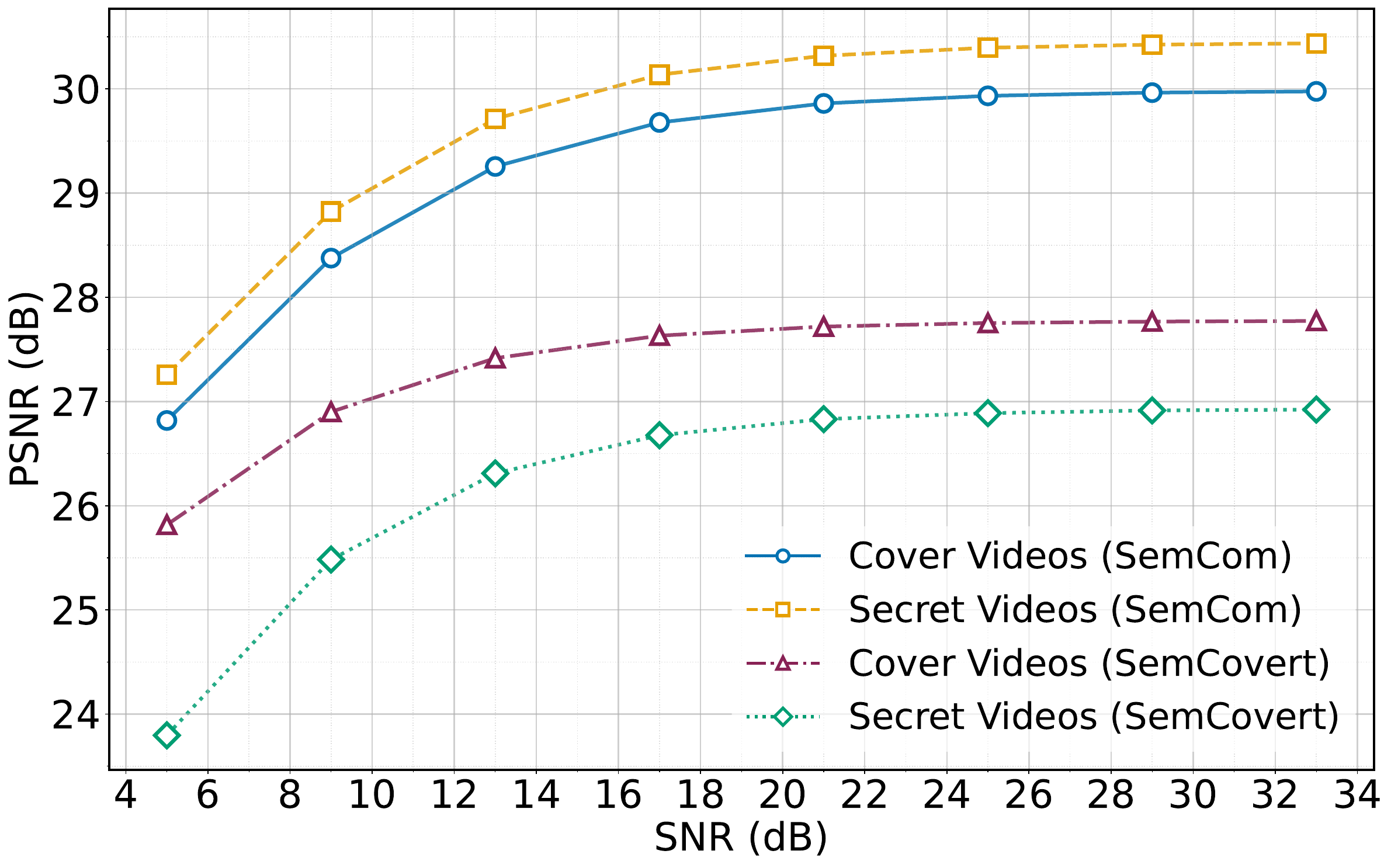}
        }
    \subfloat[SSIM comparison in DAVIS. Size: 480p.]{
        \label{fig.main_result5}
        \includegraphics[width=0.32\linewidth]{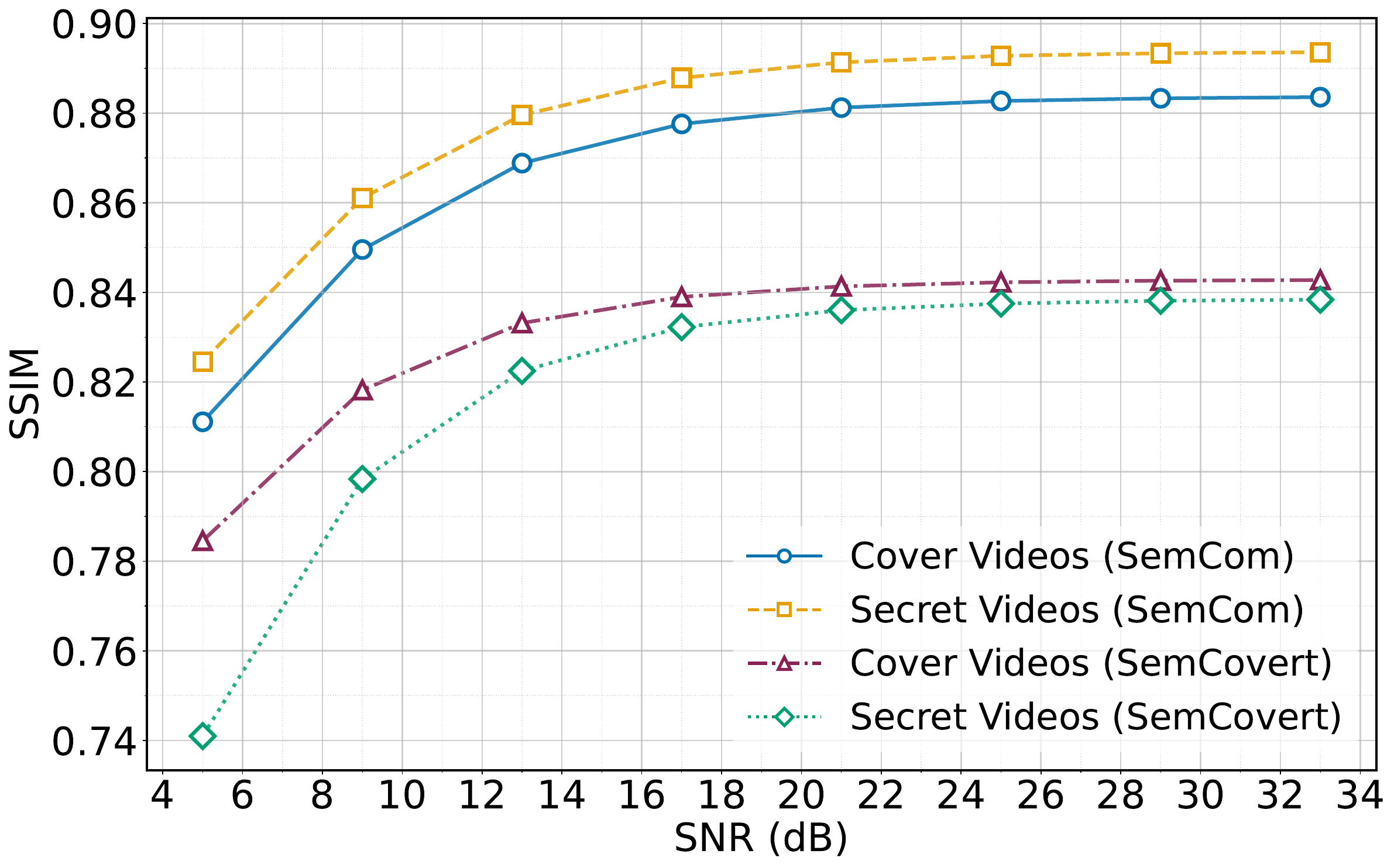}
        }
    \subfloat[FVD comparison in DAVIS. Size: 480p.]{
        \label{fig.main_result6}
        \includegraphics[width=0.32\linewidth]{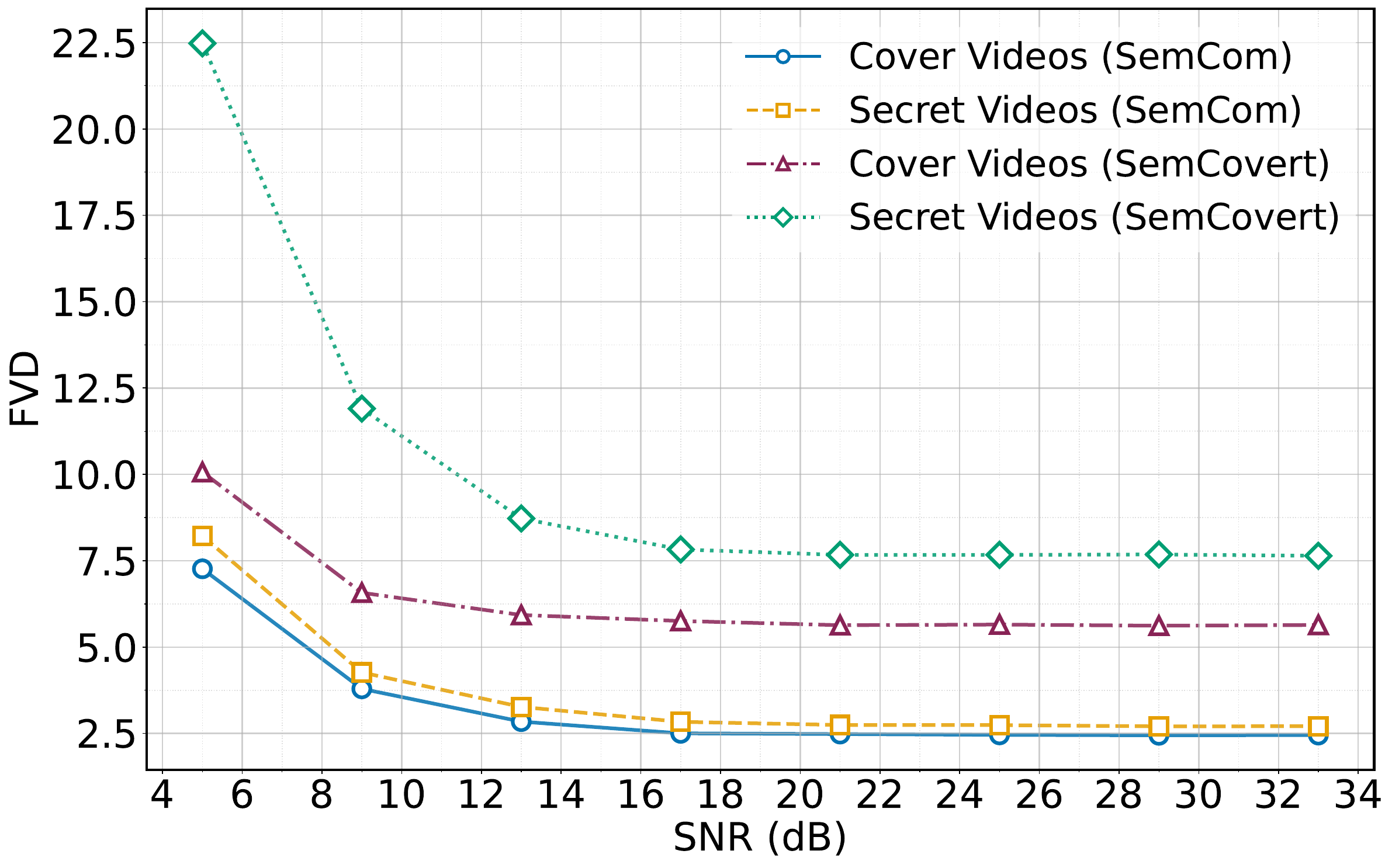}
        }
    \hfill
    \subfloat[PSNR comparison in MOT17. Size: 1080p.]{
        \label{fig.main_result7}
        \includegraphics[width=0.32\linewidth]{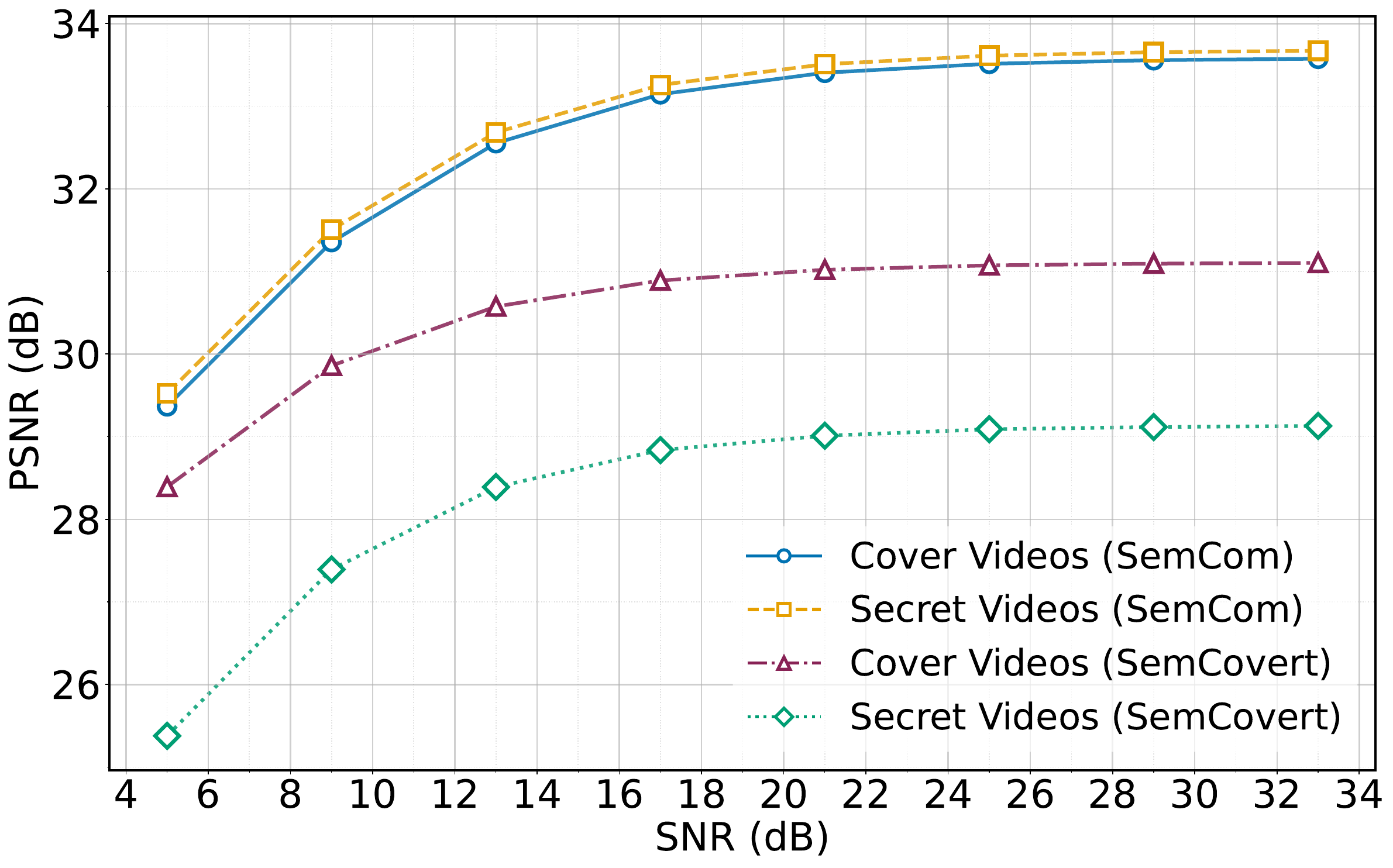}
        }
    \subfloat[SSIM comparison in MOT17. Size: 1080p.]{
        \label{fig.main_result8}
        \includegraphics[width=0.32\linewidth]{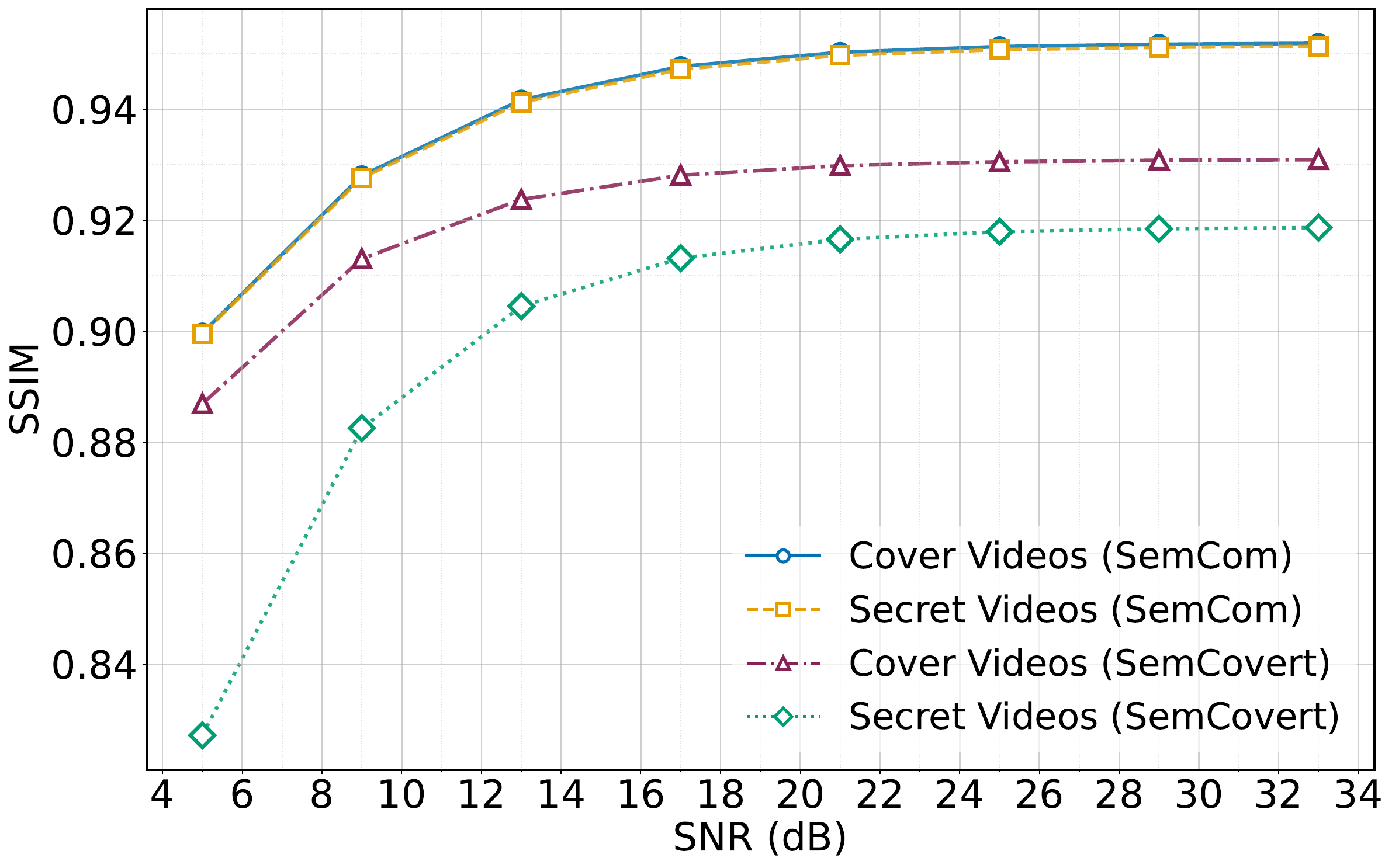}
        }
    \subfloat[FVD comparison in MOT17. Size: 1080p.]{
        \label{fig.main_result9}
        \includegraphics[width=0.32\linewidth]{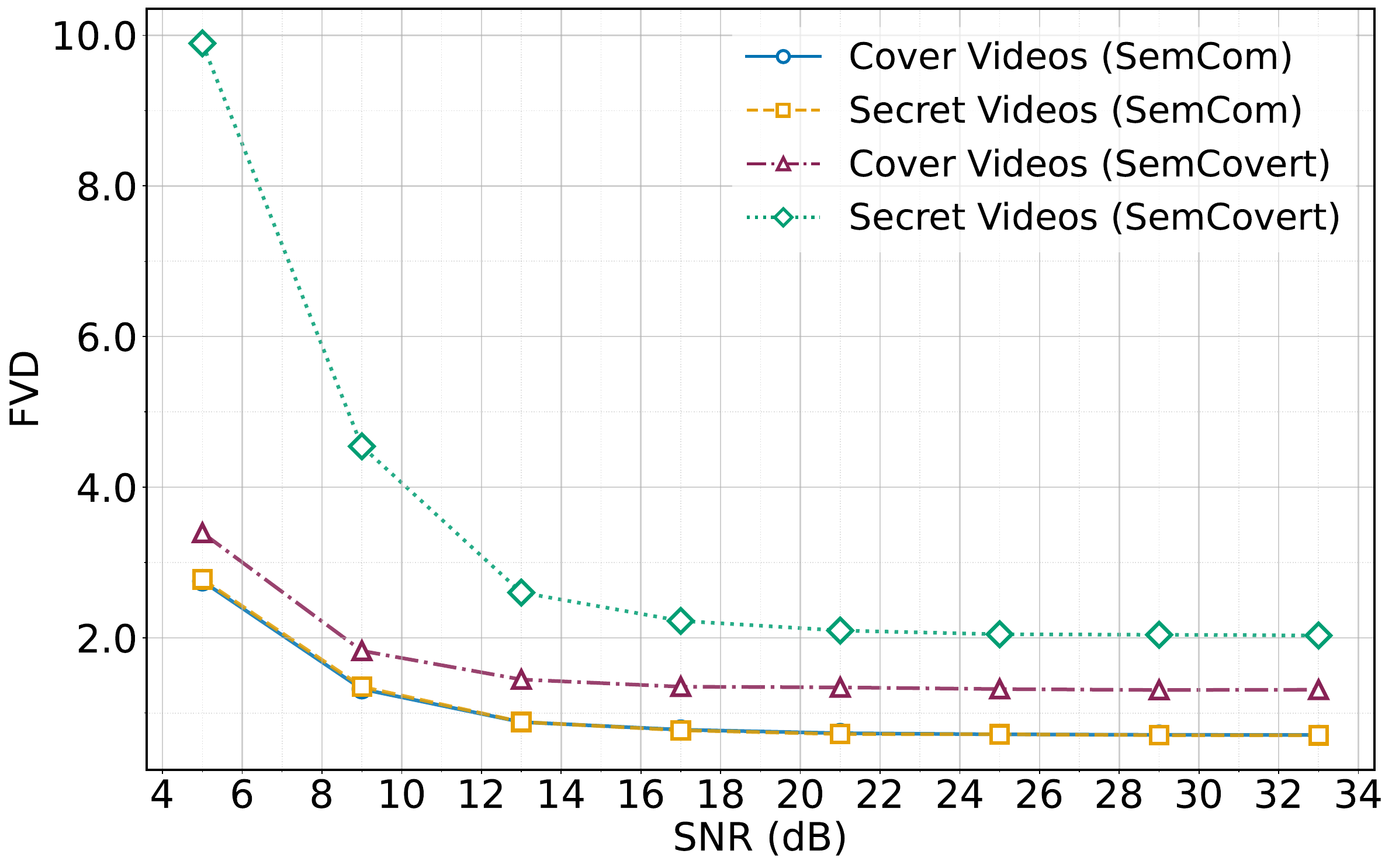}
        }
    \caption{Reconstruction quality of SemCovert and baseline methods across different resolution datasets.}
    \label{fig.main_result}
\end{figure*}

\vspace{0.8ex}
\textbf{SemCovert ensures high-quality video transmission while ensuring secure and covert transmission.}
To thoroughly assess how semantic hiding influences reconstruction quality in the context of semantic communication, we carried out a multi-faceted comparison of video outputs with and without the incorporation of SemCovert. Fig.~\ref{fig.main_result} presents a performance comparison on three datasets, leveraging PSNR, SSIM, and FVD as evaluation metrics, for videos recovered via standard video semantic communication (SemCom) and those transmitted through semantic hiding.

On the UCF101 dataset, as channel SNR increases, both cover and secret video reconstruction quality improves. When SNR exceeds $25$, the recovered original cover reaches a PSNR above $30$, while the steganographic version achieves $28$, with a consistently small PSNR gap ($\Delta \text{PSNR} < 2$). Even under low SNR, the gap remains minimal ($\Delta \text{PSNR} \approx 0.5$, $\Delta \text{SSIM} < 0.04$). Though the hidden secret video yields slightly lower quality, it still maintains high fidelity, with $\Delta \text{PSNR}$ within $4$–$5$, $\Delta \text{SSIM}$ between $0.05$–$0.1$, and $\Delta \text{FVD}$ from $1$ to $5$—gaps that further narrow under better channel conditions. These results show that secret videos can be embedded and transmitted effectively without significantly degrading the cover video quality, ensuring both fidelity and stealth.

\textbf{SemCovert demonstrates strong generalization ability.}
We tested the model trained on UCF101 on DAVIS and MOT17, which have more complex scenes and higher video resolution. The results showed that robustness was well sustained. As seen in Fig. ~\ref{fig.main_result}, on DAVIS, both videos retain high quality under high SNR ($\text{PSNR} > 25$, $\text{SSIM} > 0.84$, $\text{FVD} < 8$), indicating strong generalization. Secret video recovery under semantic-level hiding remains accurate ($\Delta \text{PSNR} < 4$, $\Delta \text{SSIM} < 0.06$, $\Delta \text{FVD} < 5$), even with more complex content and resolution.
On MOT17, performance improves further. High-SNR conditions yield sharper reconstructions for both cover and secret, with $\Delta \text{PSNR}$ within $4$, $\Delta \text{SSIM}$ below $0.03$, and $\Delta \text{FVD}$ under $2$—substantiating efficient and reliable semantic hiding.

Overall, SemCovert consistently achieves high-quality reconstruction and stable secret transmission across diverse datasets, demonstrating strong generalization, robustness, and semantic communication fidelity.

\vspace{0.8ex}
\textbf{SemCovert is capable of achieving an excellent trade-off between transmission quality and compression ratio.}
Tab.~\ref{table.exp3} reports the transmission quality and compression performance of SemCovert when transmitting cover videos under different capacity ratios. The Compression Ratio (CR) is defined as the size ratio between the compressed semantic representation and the original video data. Specifically, it is computed based on the number of channels C, temporal length T, spatial height H, and width W of the compressed representation, relative to the corresponding dimensions of the uncompressed video. As the capacity ratio increases, a larger proportion of semantic representations are used for secret embedding, resulting in lower PSNR, SSIM, and higher FVD values. In the extreme case of high capacity, the PSNR, SSIM, and FVD reach $28.53$, $0.901$, and $1.942$, respectively, showing a noticeable gap compared to the case without secret embedding (capacity ratio equals $0$).
Nevertheless, the randomized semantic hiding strategy provides a controllable mechanism for adjusting the capacity ratio, allowing SemCovert to flexibly balance transmission quality and compression efficiency. Moreover, by properly selecting the capacity ratio, SemCovert can significantly increase the difficulty of detection for potential attackers while maintaining transmission quality, as also proved by the results in Tab.~\ref{table.exp2}.

\begin{table}[!t]
    \centering
    \renewcommand{\arraystretch}{1.4}
    \caption{The transmission quality and Compression Ratio (CR) of Cover Video under different Capacity Ratios \(\frac{N}{M}\).}
    \begin{tabularx}{\linewidth}{P{2.6cm}ZZZZ}
    \thickhline
        \textbf{Capacity Ratio $\frac{N}{M}$}& \textbf{PSNR$\uparrow$}& \textbf{SSIM$\uparrow$} & \textbf{FVD$\downarrow$} & \textbf{CR$\downarrow$} \\ 
    \midrule
    \midrule
        $\mathbf{0}$ & $31.16$ & $0.931$ & $1.474$ & $0.033$  \\ 
        $\mathbf{0.2}$ & $30.36$ & $0.927$ & $1.499$ & $0.028$  \\ 
        $\mathbf{0.4}$ & $29.67$ & $0.921$ & $1.569$ & $0.024$ \\ 
        $\mathbf{0.6}$ & $29.04$ & $0.917$ & $1.778$ & $0.021$  \\ 
        $\mathbf{0.8}$ & $28.73$ & $0.911$ & $1.896$ & $0.019$  \\ 
        $\mathbf{1.0}$ & $28.53$ & $0.901$ & $1.942$ & $0.016$  \\ 
    \thickhline
    
    \label{table.exp3}
    \end{tabularx}
\end{table}

    

\vspace{0.8ex}
\textbf{SemCovert is robust against adversarial attacks.} 
We evaluate the robustness of SemCovert against adversarial attacks using three widely adopted methods: FGSM~\cite{goodfellow2014explaining}, PGD~\cite{madry2017towards}, and CW~\cite{7958570}. The adversarial objective is defined as degrading the recovery quality of the secret video as much as possible while keeping the reconstruction quality of the cover video nearly unchanged. The experimental results are summarized in Tab.~\ref{table.exp4}.
As evidenced from the results, none of the three adversarial attacks causes significant performance degradation. For both the cover and secret videos, the variation in PSNR remains within $1$, the change in SSIM is below $0.1$, and the change in FVD is also within $1$. These observations indicate that the applied adversarial perturbations fail to effectively disrupt the semantic hiding and recovery process, demonstrating that SemCovert exhibits a certain degree of robustness against typical adversarial attacks.


    

\begin{table}[!t]
    \centering
    \renewcommand{\arraystretch}{1.4}
    \caption{Secret extraction efficacy of secure transmission methods under SemCom pipeline.}
    \begin{tabularx}{\linewidth}{P{2.0cm}P{1.6cm}ZZZ}
    \thickhline
        \textbf{Types} & \textbf{Methods}& \textbf{$\Delta $PSNR}& \textbf{$\Delta $SSIM} & \textbf{$\Delta $FVD}  \\ 
    \midrule
    \midrule
        \multirow{3}{*}{Cover Videos} &
        FGSM \cite{goodfellow2014explaining} & $0.56$ & $0.02$ & $0.77$  \\ 
        & PGD \cite{madry2017towards} & $0.62$ & $0.02$ & $0.63$  \\ 
        & CW \cite{7958570} & $0.64$ & $0.03$ & $0.72$  \\ 

    \midrule
        \multirow{3}{*}{Secret Videos} &
         FGSM \cite{goodfellow2014explaining}  & $0.71$ & $0.04$ & $0.74$  \\ 
        & PGD \cite{madry2017towards} & $0.69$ & $0.07$ & $0.97$  \\ 
        & CW \cite{7958570} & $0.82$ & $0.05$ & $0.75$  \\ 
    \thickhline
    
    \label{table.exp4}
    \end{tabularx}
\end{table}


    

\vspace{0.8ex}
\textbf{SemCovert is lightweight and efficient.} 
Experiments were conducted on an RTX 4090. At 240p resolution, the Semantic Hiding Model takes $2.11$\,ms to process a single video chunk, while the Secret Semantic Extractor takes $2.46$\,ms. At 480p, the times are $5.11$\,ms and $5.44$\,ms, respectively. At 1080p, they are $29.56$\,ms and $29.86$\,ms, respectively. If conditions permit, using multi-batch inference will be faster. The latency scales smoothly with resolution, demonstrating strong adaptability. With fast inference and minimal overhead, SemCovert enables semantic hiding while seamlessly integrating into real-time semantic communication without disrupting primary tasks.

\subsection{Visualization}
\begin{figure*}[t]%
    \includegraphics[width=0.95\linewidth]{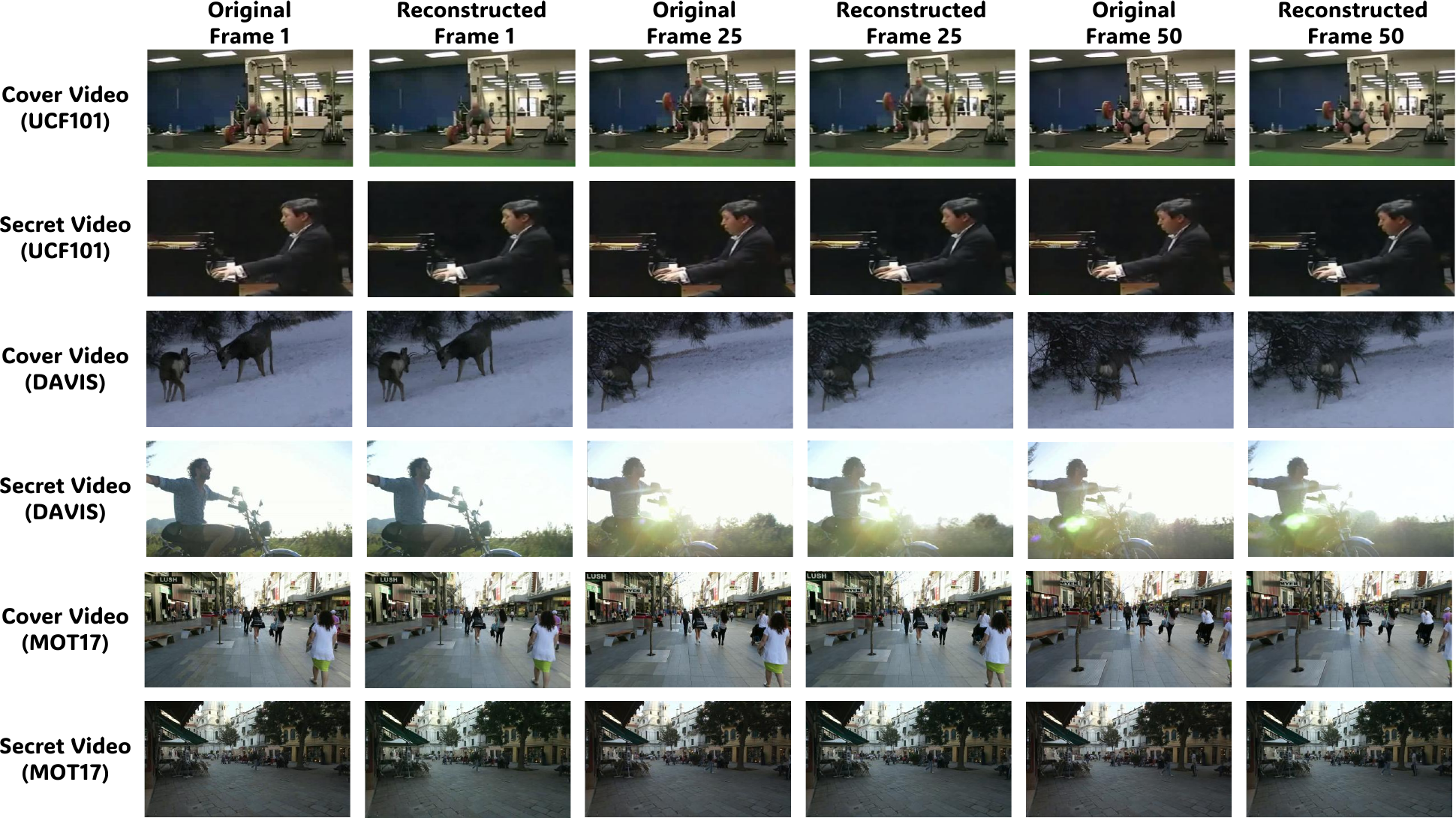}
    \caption{Visual reconstruction performance of SemCovert on cover and secret videos from multiple datasets.}
    \label{fig.visual}
\end{figure*}

To intuitively illustrate the reconstruction quality after semantic hiding and semantic communication, we present visual results in Fig.~\ref{fig.visual}. As illustrated, both the cover video and the secret video are effectively recovered, exhibiting no perceptible visual differences from the original content. We extended our evaluation to higher-resolution videos, including 480p sequences from DAVIS and 1080p sequences from MOT17, and observed that the method maintained consistently strong visual fidelity across both settings. Key information was well preserved, with no noticeable video jitter or block artifacts. These visualization results demonstrate the robustness of the Semantic Hiding Model and the Secret Semantic Extractor in SemCovert, highlighting their capability to handle videos of varying types and resolutions.

\section{Discussion}


\vspace{0.8ex}
\noindent \textbf{Why the randomized semantic hiding strategy in SemCovert makes transmission more secure and covert?}

The random semantic hiding strategy inserts secret information into only localized semantic segments, substantially reducing global perturbations and avoiding anomalies introduced by consecutive modifications. This sparse, localized approach preserves the statistical and temporal characteristics of the original sequence, thereby lowering detectability by statistical or temporal anomaly models.
Crucially, hiding positions are dynamically determined through randomized scheduling, yielding an unpredictable temporal distribution that enhances untraceability and limits the generalization capacity of detection models on unseen inputs.
Despite this randomness, SemCovert precisely identifies the hidden segments and enables high-fidelity recovery even under channel disturbances. Moreover, the reconstructed cover video accessible to potential adversaries remains markedly different from the hidden content across pixel, frequency, and semantic domains, making reverse inference highly challenging.
Finally, longer cover videos provide a broader temporal hiding space, enabling lower hiding density per frame and further improving concealment with only a moderate reduction in capacity.

\vspace{0.8ex}
\noindent \textbf{What other benefits does semantic-level hiding offer in video transmission?}

This work verifies the effectiveness of the semantic-level hiding strategy in video transmission through reconstruction quality evaluation and visual results. The proposed method fully leverages the redundancy of semantic information and the strong modeling capability of deep neural networks to achieve information embedding without introducing additional transmission overhead. Experimental results show that high-fidelity video reconstruction and semantic recovery can still be achieved even under poor channel conditions.
Notably, the method maintains stable one-to-one embedding at compression rates exceeding 30 and demonstrates strong adaptability across different video resolutions, indicating high robustness and transferability. This performance surpasses the limitations of traditional secure transmission methods constrained by the Shannon limit, highlighting the unique advantages of semantic communication in complex transmission scenarios.
This study focuses on the security of semantic video communication, while leaving ample room for future exploration in this area.

\section{Related Work}
\subsection{Video Semantic Communication}

SOTA research in video SemCom has addressed multiple aspects of the field \cite{10539255, 10685066, 10740049}. Expanding on traditional SemCom, \cite{9955991} reduces bandwidth and enhances error resilience by utilizing keypoint transmission and a novel semantic error detection mechanism. To improve efficiency in noisy channels, \cite{10464666} proposed MDVSC, which enhances video transmission through semantic feature extraction, entropy-based variable-length coding, and a learnable architecture. \cite{10570803} introduced an end-to-end SemCom for large-scale communication, focusing on key objects and spatiotemporal correlations using convolutional motion estimation, contextual extraction, and entropy coding. To address latency and accuracy challenges in fixed bitrate transmission, \cite{10405124} developed an adaptive bitrate SemCom system that adjusts based on network conditions, using a Swin Transformer-based codec and Actor-Critic-based ABR algorithm. \cite{10622487} presented a synchronous SemCom system for video, aligning semantic and temporal multimodal data through 3D Morphable Model coefficients and visual-guided speech synthesis, optimizing overhead and quality.

\subsection{Secure Transmission Techniques}

\subsubsection{Steganography}
This is a secure communication technique that embeds hidden information within video carriers while preserving visual quality \cite{LIU2019238}. Current mainstream approaches employ three primary methodologies: \textit{\textbf{i}}) Spatial-domain methods directly modify pixel values in video frames and offer simplicity and low computational overhead \cite{10.1007/978-981-10-6890-4_67}.
\textit{\textbf{ii}}) Frequency-domain techniques leverage transforms like DCT to embed data in frequency coefficients, significantly enhancing resistance to compression and interference \cite{6615345}.
\textit{\textbf{iii}}) Temporal-domain approaches exploit video sequence characteristics by utilizing inter-frame motion features for data concealment \cite{10.1145/3323873.3325011}.
Recent advances in deep learning have enabled neural network-based steganographic algorithms to demonstrate remarkable adaptability, dynamically adjusting embedding strategies according to video content \cite{10.1007/978-981-19-1677-9_64}. Also, breakthroughs in dynamic video steganography was achieved via intelligently recognizing moving objects in scene variations, yielding more covert and practical detail hiding \cite{DBLP:phd/ethos/Peng20}.

\subsubsection{Encryption}
Modern video encryption techniques primarily employ two paradigms: \textit{\textbf{i}}) conventional cryptographic methods and \textit{\textbf{ii}}) content-adaptive approaches \cite{LIU20103,WU2021107911,8239690}. Traditional algorithms like AES and RSA provide robust security but face computational bottlenecks in real-time scenarios. Contrastively, contemporary content-based methods (\textit{e.g.}, selective frame encryption and dynamic block ciphering) optimize the security-efficiency trade-off by leveraging video structural attributes. The advent of AI-driven solutions has further revolutionized the field, promoting intelligent content analysis for adaptive encryption strategies.

\section{Conclusion}
In this work, we propose SemCovert, a novel semantic-level covert communication framework designed to address privacy vulnerabilities in video semantic communication systems. Unlike conventional encryption and steganography, which are vulnerable to semantic transformations and lack robustness under long-term temporal modeling, SemCovert integrates a co-designed semantic hider and secret extractor directly into the communication pipeline. This allows authorized receivers to reliably recover hidden information while maintaining imperceptibility to unauthorized observers. To further enhance security, we introduce a randomized semantic hiding strategy that disrupts embedding patterns, effectively resisting distributional analysis and reverse engineering. Experimental results demonstrate that SemCovert achieves high stealthiness with negligible impact on video quality, validating its effectiveness in enabling secure and covert video transmission.

\balance
\bibliographystyle{ieeetr}
\bibliography{ref}

@ARTICLE{9955991,
  author={Jiang, Peiwen and Wen, Chao-Kai and Jin, Shi and Li, Geoffrey Ye},
  journal={IEEE J. Sel. Areas Commun.}, 
  title={Wireless Semantic Communications for Video Conferencing}, 
  year={2023},
  volume={41},
  number={1},
  pages={230-244},
}

@INPROCEEDINGS{10464666,
  author={Bao, Zhicheng and Liang, Haotai and Dong, Chen and Xu, Xiaodong and Liu, Geng},
  booktitle={Proc. IEEE Glob. Commun. Conf. (Globecom)}, 
  title={MDVSC—Wireless Model Division Video Semantic Communication for 6G}, 
  year={2023},
  volume={},
  number={},
  pages={1572-1578},
}

@INPROCEEDINGS{10570803,
  author={Li, Haopeng and Tong, Haonan and Wang, Sihua and Yang, Nuocheng and Yang, Zhaohui and Yin, Changchuan},
  booktitle={Proc. IEEE Wireless Commun. Networking Conf. (WCNC)}, 
  title={Video Semantic Communication with Major Object Extraction and Contextual Video Encoding}, 
  year={2024},
  volume={},
  number={},
  pages={1-6},
}

@INPROCEEDINGS{10405124,
  author={Gong, Wentao and Tong, Haonan and Wang, Sihua and Yang, Zhaohui and He, Xinxin and Yin, Changchuan},
  booktitle={Proc. Int. Conf. Wireless Commun. Signal Process. (WCSP)}, 
  title={Adaptive Bitrate Video Semantic Communication over Wireless Networks}, 
  year={2023},
  volume={},
  number={},
  pages={122-127},
}

@INPROCEEDINGS{10622487,
  author={Tian, Yun and Ying, Jingkai and Qin, Zhijin and Jin, Ye and Tao, Xiaoming},
  booktitle={Proc. IEEE Int. Conf. Commun. (ICC)}, 
  title={Synchronous Semantic Communications for Video and Speech}, 
  year={2024},
  volume={},
  number={},
  pages={3396-3401},
}

@ARTICLE{10685066,
  author={Bao, Zhicheng and Liang, Haotai and Dong, Chen and Li, Cong and Xu, Xiaodong and Zhang, Ping},
  journal={IEEE Internet Things J.}, 
  title={MDVSC—Efficient Wireless Model Division Video Semantic Communication}, 
  year={2025},
  volume={12},
  number={2},
  pages={1109-1124},
}

@ARTICLE{10740049,
  author={Tong, Haonan and Li, Haopeng and Du, Hongyang and Yang, Zhaohui and Yin, Changchuan and Niyato, Dusit},
  journal={IEEE Wirel. Commun. Lett.}, 
  title={Multimodal Semantic Communication for Generative Audio-Driven Video Conferencing}, 
  year={2025},
  volume={14},
  number={1},
  pages={93-97},
}

@ARTICLE{10539255,
  author={Zhang, Bowen and Qin, Zhijin and Li, Geoffrey Ye},
  journal={IEEE J. Sel. Top. Signal Process.}, 
  title={Compression Ratio Learning and Semantic Communications for Video Imaging}, 
  year={2024},
  volume={18},
  number={3},
  pages={312-324},
}

@ARTICLE{10974507,
  author={Du, Qiyuan and Duan, Yiping and Yang, Qianqian and Tao, Xiaoming and Debbah, Mérouane},
  journal={IEEE J. Sel. Areas Commun.}, 
  title={Object-Attribute-Relation Representation-Based Video Semantic Communication}, 
  year={2025},
  volume={43},
  number={7},
  pages={2446-2461},
}

@article{soomro2012ucf101,
  title={Ucf101: A dataset of 101 human actions classes from videos in the wild},
  author={Soomro, Khurram and Zamir, Amir Roshan and Shah, Mubarak},
  journal={arXiv preprint arXiv:1212.0402},
  year={2012}
}

@inproceedings{Perazzi2016,
  author = {F. Perazzi and J. Pont-Tuset and B. McWilliams and L. {Van Gool} and M. Gross and A. Sorkine-Hornung},
  title = {A Benchmark Dataset and Evaluation Methodology for Video Object Segmentation},
  booktitle = {Proc. IEEE/CVF Comput. Vis. Pattern Recognit. (CVPR)},
  year = {2016}
}

@article{milan2016mot16,
  title={MOT16: A benchmark for multi-object tracking},
  author={Milan, Anton and Leal-Taix{\'e}, Laura and Reid, Ian and Roth, Stefan and Schindler, Konrad},
  journal={arXiv preprint arXiv:1603.00831},
  year={2016}
}

@inproceedings{xie2018rethinking,
  title={Rethinking spatiotemporal feature learning: Speed-accuracy trade-offs in video classification},
  author={Xie, Saining and Sun, Chen and Huang, Jonathan and Tu, Zhuowen and Murphy, Kevin},
  booktitle={Proc. Eur. Conf. Comput. Vis. (ECCV)},
  pages={305--321},
  year={2018}
}

@article{LIU2019238,
title = {Video steganography: A review},
journal = {Neurocomputing},
volume = {335},
pages = {238-250},
year = {2019},
issn = {0925-2312},
author = {Yunxia Liu and Shuyang Liu and Yonghao Wang and Hongguo Zhao and Si Liu},
}

@InProceedings{10.1007/978-981-10-6890-4_67,
author="Dalal, Mukesh
and Juneja, Mamta",

title="Video Steganography Techniques in Spatial Domain---A Survey",
booktitle="Proc. Int. Conf. Comput. Commun. Syst. (ICCCS)",
year="2018",
pages="705--711",
}

@inproceedings{10.1145/3323873.3325011,
author = {Weng, Xinyu and Li, Yongzhi and Chi, Lu and Mu, Yadong},
title = {High-Capacity Convolutional Video Steganography with Temporal Residual Modeling},
year = {2019},
booktitle = {Proc. Int. Conf. Multimed. Retr. (ICMR)},
pages = {87–95},
}

@INPROCEEDINGS{6615345,
  author={Niu Ke and Zhong Weidong},
  booktitle={Proc. IEEE Int. Conf. Softw. Eng. Serv. Sci. (ICSESS)}, 
  title={A video steganography scheme based on H.264 bitstreams replaced}, 
  year={2013},
  volume={},
  number={},
  pages={447-450},
}

@InProceedings{10.1007/978-981-19-1677-9_64,
author="Kumar, P. Sathish
and Fathima, K.
and Karthik, B.
and Kumar, S. Siva
and Sowmya, B.
and Ghosh, Ankush",
title="Studies on Steganography Images and Videos Using Deep Learning Techniques",
booktitle="Innov. Electr. Electron. Eng.",
year="2022",
pages="707--733",
}

@phdthesis{DBLP:phd/ethos/Peng20,
  author       = {Jinghui Peng},
  title        = {Secure covert communications over streaming media using dynamic steganography},
  school       = {University of West London, Ealing, {UK}},
  year         = {2020},
}

@article{LIU20103,
title = {A survey of video encryption algorithms},
journal = {Comput. Secur.},
volume = {29},
number = {1},
pages = {3-15},
year = {2010},
author = {Fuwen Liu and Hartmut Koenig},
}

@article{WU2021107911,
title = {Content-adaptive image encryption with partial unwinding decomposition},
journal = {Signal Proc.},
volume = {181},
pages = {107911},
year = {2021},
author = {Yongfei Wu and Liming Zhang and Tao Qian and Xilin Liu and Qiwei Xie},
}

@ARTICLE{8239690,
  author={Li, Peiya and Lo, Kwok-Tung},
  journal={IEEE Trans. Multim.}, 
  title={A Content-Adaptive Joint Image Compression and Encryption Scheme}, 
  year={2018},
  volume={20},
  number={8},
  pages={1960-1972},
  doi={10.1109/TMM.2017.2786860}
}

@article{simonyan2014very,
  title={Very deep convolutional networks for large-scale image recognition},
  author={Simonyan, Karen and Zisserman, Andrew},
  journal={arXiv preprint arXiv:1409.1556},
  year={2014}
}

@article{wan2025,
      title={Wan: Open and Advanced Large-Scale Video Generative Models}, 
      author={Team Wan and Ang Wang and Baole Ai and Bin Wen and others},
      journal = {arXiv preprint arXiv:2503.20314},
      year={2025}
}

@misc{kingma2013auto,
  title={Auto-encoding variational bayes},
  author={Kingma, Diederik P and Welling, Max and others}
}

@inproceedings{tran2018closer,
  title={A closer look at spatiotemporal convolutions for action recognition},
  author={Tran, Du and Wang, Heng and Torresani, Lorenzo and Ray, Jamie and LeCun, Yann and Paluri, Manohar},
  booktitle={Proc. IEEE/CVF Comput. Vis. Pattern Recognit. (CVPR)},
  pages={6450--6459},
  year={2018}
}

@inproceedings{liu2022video,
  title={Video swin transformer},
  author={Liu, Ze and Ning, Jia and Cao, Yue and Wei, Yixuan and Zhang, Zheng and Lin, Stephen and Hu, Han},
  booktitle={Proc. IEEE/CVF Comput. Vis. Pattern Recognit. (CVPR)},
  pages={3202--3211},
  year={2022}
}

@misc{zhang2019rootmeansquarelayer,
      title={Root Mean Square Layer Normalization}, 
      author={Biao Zhang and Rico Sennrich},
      year={2019},
      eprint={1910.07467},
      archivePrefix={arXiv},
      primaryClass={cs.LG},
      url={https://arxiv.org/abs/1910.07467}, 
}

@article{telli2024new,
  title={A new approach to video steganography models with 3D deep CNN autoencoders},
  author={Telli, Mounir and Othmani, Mohamed and Ltifi, Hela},
  journal={Multimed. Tools Appl.},
  volume={83},
  number={17},
  pages={51423--51439},
  year={2024},
  publisher={Springer}
}

@article{xu2022end,
  title={An End-to-End Robust Video Steganography Model Based on a Multi-Scale Neural Network},
  author={Xu, Shutong and Li, Zhaohong and Zhang, Zhenzhen and Liu, Junhui},
  journal={Electron.},
  volume={11},
  number={24},
  pages={4102},
  year={2022},
  publisher={MDPI}
}

@article{ShengFLCWS25,
  author={Qingxin Sheng and Chong Fu and Zhaonan Lin and others},
  title={Content-Aware Tunable Selective Encryption for HEVC Using Sine-Modular Chaotification Model},
  year={2025},
  cdate={1735689600000},
  journal={IEEE Trans. Multim.},
  volume={27},
  pages={41-55},
  url={https://doi.org/10.1109/TMM.2024.3521724}
}

@article{abomhara2010overview,
  title={An overview of video encryption techniques},
  author={Abomhara, Mohamed and Zakaria, Omar and Khalifa, Othman O},
  journal={Int. J. Comput. Theory Eng.},
  volume={2},
  number={1},
  pages={1793--8201},
  year={2010},
  publisher={IACSIT Press}
}

@article{yang2024secure,
  title={Secure semantic communications: Fundamentals and challenges},
  author={Yang, Zhaohui and Chen, Mingzhe and Li, Gaolei and Yang, Yang and Zhang, Zhaoyang},
  journal={IEEE Netw.},
  volume={38},
  number={6},
  pages={513--520},
  year={2024},
  publisher={IEEE}
}

@article{guo2025videoqa,
  title={VideoQA-SC: Adaptive semantic communication for video question answering},
  author={Guo, Jiangyuan and Chen, Wei and Sun, Yuxuan and Xu, Jialong and Ai, Bo},
  journal={IEEE J. Sel. Areas Commun.},
  year={2025},
  publisher={IEEE}
}

@article{huo2024image,
  title={Image semantic steganography: A way to hide information in semantic communication},
  author={Huo, Yanhao and Xiang, Shijun and Luo, Xiangyang and Zhang, Xinpeng},
  journal={IEEE Trans. Circuits Syst. Video Technol.},
  year={2024},
  publisher={IEEE}
}

@article{kunhoth2023video,
  title={Video steganography: recent advances and challenges},
  author={Kunhoth, Jayakanth and Subramanian and others},
  journal={Multimed. Tools Appl.},
  volume={82},
  number={27},
  pages={41943--41985},
  year={2023},
  publisher={Springer}
}

@InProceedings{lvni,
    author    = {Mou, Chong and Xu, Youmin and Song, Jiechong and Zhao, Chen and Ghanem, Bernard and Zhang, Jian},
    title     = {Large-Capacity and Flexible Video Steganography via Invertible Neural Network},
    booktitle = {Proc. IEEE/CVF Comput. Vis. Pattern Recognit. (CVPR)},
    month     = {June},
    year      = {2023},
    pages     = {22606-22615}
}

@inproceedings{mao2024covert,
  title={From covert hiding to visual editing: robust generative video steganography},
  author={Mao, Xueying and Hu, Xiaoxiao and Peng, Wanli and Gan, Zhenliang and Qian, Zhenxing and Zhang, Xinpeng and Li, Sheng},
  booktitle={ACM Int. Conf. Multimedia},
  pages={2757--2765},
  year={2024}
}

@INPROCEEDINGS{9878477,
  author={Xu, Youmin and Mou, Chong and Hu, Yujie and Xie, Jingfen and Zhang, Jian},
  booktitle={Proc. IEEE/CVF Comput. Vis. Pattern Recognit. (CVPR)}, 
  title={Robust Invertible Image Steganography}, 
  year={2022},
  volume={},
  number={},
  pages={7865-7874},
  doi={10.1109/CVPR52688.2022.00772}}

@article{goodfellow2014explaining,
  title={Explaining and harnessing adversarial examples},
  author={Goodfellow, Ian J and Shlens, Jonathon and Szegedy, Christian},
  journal={arXiv preprint arXiv:1412.6572},
  year={2014}
}

@article{madry2017towards,
  title={Towards deep learning models resistant to adversarial attacks},
  author={Madry, Aleksander and Makelov, Aleksandar and Schmidt, Ludwig and Tsipras, Dimitris and Vladu, Adrian},
  journal={arXiv preprint arXiv:1706.06083},
  year={2017}
}

@INPROCEEDINGS{7958570,
  author={Carlini, Nicholas and Wagner, David},
  booktitle={Proc. IEEE Symp. Secur. Priv. (SP)}, 
  title={Towards Evaluating the Robustness of Neural Networks}, 
  year={2017},
  volume={},
  number={},
  pages={39-57},
  doi={10.1109/SP.2017.49}}

\end{document}